\documentclass[twocolumn,english,aps, prx, eqsecnum]{revtex4-2}
\usepackage[T1]{fontenc}
\usepackage[latin9]{inputenc}
\usepackage{color}
\usepackage{bm}
\usepackage{amsmath}
\usepackage{amssymb}
\usepackage{graphicx}
\usepackage{babel}
\begin{document}
\title{The coherent Ising machine with quantum feedback: the total and conditional
master equation methods}
\author{Simon Kiesewetter}
\author{Peter D Drummond}
\affiliation{Centre for Quantum Science and Technology Theory, Swinburne University
of Technology, Melbourne 3122, Australia}
\begin{abstract}
We give a detailed theoretical derivation of the master equation for
the coherent Ising machine. This is a quantum computational network
with feedback, that approximately solves \emph{NP} hard combinatoric
problems, including the traveling salesman problem and various extensions
and analogs. There are two possible types of master equation, either
conditional on the feedback current or unconditional. We show that
both types can be accurately simulated in a scalable way using stochastic
equations in the positive-P phase-space representation. This depends
on the nonlinearity present, and we use parameter values that are
typical of current experiments. While the two approaches are in excellent
agreement, they are not equivalent with regard to efficiency. We find
that unconditional simulation has much greater efficiency, and is
more scalable to large sizes. This is a case where too much knowledge
is a dangerous thing. Conditioning the simulations on the feedback
current is not essential to determining the success probability, but
it greatly increases the computational complexity. To illustrate the
speed improvements obtained with the unconditional approach, we carry
out full quantum simulations of the  master equation with up to $1000$
nodes.
\end{abstract}
\maketitle

\section{Introduction}

The coherent Ising machine (CIM) is a type of computational device
which operates in a fundamentally different way to both classical
and gate-based quantum computers. It has been known for a while that
there is a wide variety of computationally challenging (NP-complete
or NP-hard) \citep{lucas2014ising,Glover2018ATO} problems that can
be mapped onto the Ising model. This is a simple model consisting
of binary variables, usually identified with spins in a magnetic material
which interact both locally and non-locally to give a Hamiltonian
model whose ground state is the solution to the computational problem.
Originally, the model corresponded to spins interacting with an external
magnetic field and with each other through spin-spin coupling. Since
a true Ising model using physical spins is difficult to manipulate
experimentally, the CIM aims to simulate it, using continuous variables
in a non-equilibrium setup whose steady-state closely resembles the
Ising model. The largest experiments of this type \citep{honjo2021100}
use a measurement-feedback strategy \citep{McMahon2016}.

We compare two different techniques of simulating the measurement-feedback
CIM using the positive-P phase space representation \citep{Drummond_generalizedP1980,Drummond_OpActa1981},
which is an exact mapping of quantum dynamics to stochastic equations.
Phase space simulation using the positive-P representation provides
a convenient, scalable way to simulate the nonlinear system dynamics
of some types of complex, dissipative quantum systems without the
need to make approximations. Except for rare cases with very low losses,
where non-vanishing boundary terms are present, this method gives
quantitative predictions. We show that there is no need to make any
approximations of the system equations \citep{Gilchrist_PRA1997,Schack1991,Smith1989}.
However, simulating the system dynamics is complicated by the homodyne
measurement used for feedback, which causes a partial collapse of
the system wave-function according to the measurement outcome.

The measurement outcome is partly determined by quantum noise at the
measurement site. As a result, the quantum dynamics follows a conditional
master equation due to the noisy outcome of the measurement feedback.
Here we derive the multi-mode conditional master equation as a stochastic
equation in the Stratonovich calculus. The operator associated with
the wave-function collapse leads to terms which do not correspond
to a conventional Fokker-Planck equation in a phase-space representation,
and a weighted simulation is required \citep{hush2009scalable}. It
is also possible to consider an average over the feedback, giving
an unconditional master equation. Both types of equation can be exactly
simulated with the positive-P phase-space method, and we show that
they lead to identical success rate predictions.

The original gedanken-experiment \citep{Utsunomiya2011,Takata2012,wang2013coherent}
used laser pulses impinging on multiple degenerate parametric oscillators
(DPO) \citep{drummond1980non,Drummond_OpActa1981}, realized by a
nonlinear medium in an optical cavity. At a certain pump strength,
each DPO becomes a bistable system, with quantum states that are associated
with the binary variables of the Ising model, and can be coupled to
each other. Hence, an ideal DPO-based CIM is a true quantum system,
with transient states that are like a Schr\"odinger Cat state of
form $\left|\alpha\right\rangle +\left|-\alpha\right\rangle $ \citep{schrodinger_1935,Wolinsky_PRL1988,Krippner_PRA1994},
when losses are very low. Therefore it has the potential to be subject
to quantum enhancement, which may contribute to steering the system
into the desired steady-state, approximately equivalent to an Ising
ground-state. The significance of such effects is still subject to
investigation. There are also other types of realization of the CIM
via electronic or digital circuits that simulate the dynamics in a
classical regime \citep{bohm2019poor}. 

In the first practical realizations of the CIM the DPO itself was
a localized pulse stored in an optical fiber loop \citep{Marandi_CIM_Nature2014,Takata2016}.
Effective spin-spin interactions are obtained using an optical delay-line
(ODL-CIM) that redirects part of the time-delayed signal back into
the fiber loop, allowing different pulses to interact. While this
architecture has many advantages, its principal disadvantage is that
it is difficult to scale up to include large numbers of spins. Due
to the close similarity of superconducting and optical parametric
amplifiers, it may be feasible to realize this type of device in a
superconducting waveguide. Much stronger quantum effects are known
in such cases \citep{wang2016schrodinger}, and quantum tunneling
is possible \citep{Sun_NJP2019,Sun_PRA2019,teh2020dynamics}. 

A different version of the CIM, commonly called a measurement-feedback
or MFB-CIM, was developed a few years later \citep{honjo2021100,McMahon2016,shoji2017quantum,yamamura2017quantum,inagaki2016}.
Here, the signal state is observed via a homodyne detector and the
feedback strength is calculated electronically based on the measurement.
A feedback signal is then generated from the pump pulse and fed back
into the loop after a variable time-delay. This architecture has the
great advantage of being very well suited for the simulation of systems
of a large number of Ising spins. It has been demonstrated most impressively
in a recent experiment of a measurement-feedback type CIM involving
100,000 spins \citep{honjo2021100}.

Phase-space approaches have proved the only practical, scalable way
to treat large quantum networks. These are based on earlier multi-mode
quantum field simulations \citep{carter1987squeezing,raymer1991limits,drummond1993simulation},
and have already been used to analyse Gaussian boson sampling quantum
computers \citet{drummond2022simulating}. Equations based on an approximate
phase-space approach are known for an ODL-CIM \citep{wang2013coherent,maruo2016truncated}
and for an MFB-CIM architecture \citep{McMahon2016}. These use a
modified Wigner representation \citep{Wigner_PhysRev1932}, which
truncate third and higher order derivatives in the corresponding Fokker-Planck
equation. Exact positive-P equations of motion \citep{Drummond_generalizedP1980}
that do not require truncation are given both for the ODL \citep{Takata2015,maruo2016truncated}
and for the MFB-type CIM\citep{inui2020noise}. A scheme for weighted
phase-space simulations involving the conditional master equation
of a MFB-type CIM is known \citep{shoji2017quantum}. Discrete-time
descriptions of the MFB-type CIM have been published \citep{yamamura2017quantum,Ng2022PhysRevResearch.4.013009},
which use a simplified Gaussian phase-space representation \citep{corney2003gaussian}.

The term scalable refers here to the polynomial-time solution of the
CIM simulations for the given parameters and feedback method. There
is no evidence of sampling error limitations, but the observed efficient
sampling may not hold for stronger couplings, or different feedback
regimes. We do not claim that our method can accurately solve NP-hard
problems in a polynomial time, which is generally regarded as impossible
on a digital computer. However, these simulations provide a useful
way to quantitatively understand the physics and expected performance
of this quantum technology. Approximate but fast quantum hardware
solutions of these types of problem can be extremely useful in practical
applications. There can still be an experimental \textquotedbl quantum
advantage\textquotedbl , if classical polynomial time simulation
is slower than experiment. 

In this article, after reviewing the topic of quantum measurement-feedback
systems in general, we present two ways in which the the system quantum
dynamics can be simulated using exact phase-space techniques. These
correspond to the conditional and unconditional master equations approaches.
Full conditional simulation leads to an ensemble of weighted trajectories
through which the quantum master equation conditioned on the feedback
currents can be simulated \citep{hush2009scalable}. It is a relatively
complex method due to the fact that it requires a careful rebalancing
of the weight distribution to prevent numerical instabilities from
exponential growth in the weights. Alternatively, the full unconditional
master equation can be treated using unweighted stochastic trajectories,
which yields ensemble averages of quadrature measurements.

We compare the simulation outcomes and performance of the two methods.
They agree with each other extremely well in modeling success rates
of the feedback CIM. From a computational point of view, we find that
the unconditional method is greatly preferred. As it requires computing
and rebalancing weights, the conditional algorithm is more complex.
This approach also requires orders of magnitude more stochastic trajectories
to give accurate predictions. The large speed improvement in unconditional
simulations is especially important in light of the  large size of
recent measurement-feedback type CIM experiments.

\section{The Coherent Ising machine}

\subsection{The Ising model\label{subsec:The-Ising-model}}

The Ising model was formulated almost a century ago \citep{Lenz1920,Ising1925}
to model ferromagnetism and related phenomena. It is a very simple
theory, consisting of discrete variables $\sigma_{i}$, indicating
the nuclei's magnetic spins. These are oriented either ``up'' or
``down'', corresponding to $\sigma=\pm1$. The spins now interact
with each other through spin-spin interaction and with an external
magnetic field. The Ising model Hamiltonian is
\begin{eqnarray}
H & = & -\sum_{i,j}J_{ij}\sigma_{i}\sigma_{j}-\sum_{i}h_{i}\sigma_{i}\,,
\end{eqnarray}
where $\mathbf{J}$ is the coupling matrix and $h$ is proportional
to the possibly inhomogeneous magnetic field strength.

Apart from its usefulness in explaining ferromagnetism, the Ising
model has another interesting feature: a wide variety of computationally
challenging (NP-complete or NP-hard) problems can be mapped onto it
via changing the coupling matrix and investigating the corresponding
ground state. As an example, consider the so-called Max-Cut problem.
The problem statement is as follows:

Given an undirected graph $G=\left(V,E\right)$ where $V=\left\{ v_{i}\right\} $
is the set of vertices and $E=\left\{ e_{i}\right\} $ is the set
of edges and a weight function $w:E\rightarrow\mathbb{R}^{+}$, find
the bipartition (cut) into sets $V=U\uplus W$ with the highest sum
of ``weights along the cut line'', that is, maximize $f\equiv\sum_{i}w\left(e_{i}\right)$,
where $e_{i}=\left\{ u,w\right\} ,u\in U,w\in W$.

The way to map this problem onto the Ising model is to identify each
vertex with a certain spin. The interaction matrix is set to the negative
of the weights between the nodes ($0$ if there is no edge) and the
external magnetic field is set to $0$. The spin states then indicate
whether a vertex belongs to set $U$ or $W$. Upon inspecting the
system Hamiltonian, one finds that
\begin{eqnarray}
H & = & -\sum_{i,j}J_{ij}+2\sum_{\left(i,j\right)\in\Delta}J_{ij}\nonumber \\
 & = & C+2\sum_{\left(i,j\right)\in\Delta}J_{ij}\,,
\end{eqnarray}

where $\Delta$ is the set of indices $\left(i,j\right)$ such that
$v_{i}\in U,v_{j}\in W$ or vice versa. Since the total sum of weights
$C=-\sum_{i,j}J_{ij}$ is fixed, minimizing $H$ will maximize the
sum of ``cut'' weights (since $J_{ij}\le0$ by construction).

Due to the absence of Zeeman terms, the mapping between the Max-Cut
problem and the Ising model is possibly the most natural and well-known
one, however a plethora of other interesting and computationally challenging
ones can be mapped to the Ising model in a similar well.

\begin{figure}
\includegraphics[width=0.8\columnwidth]{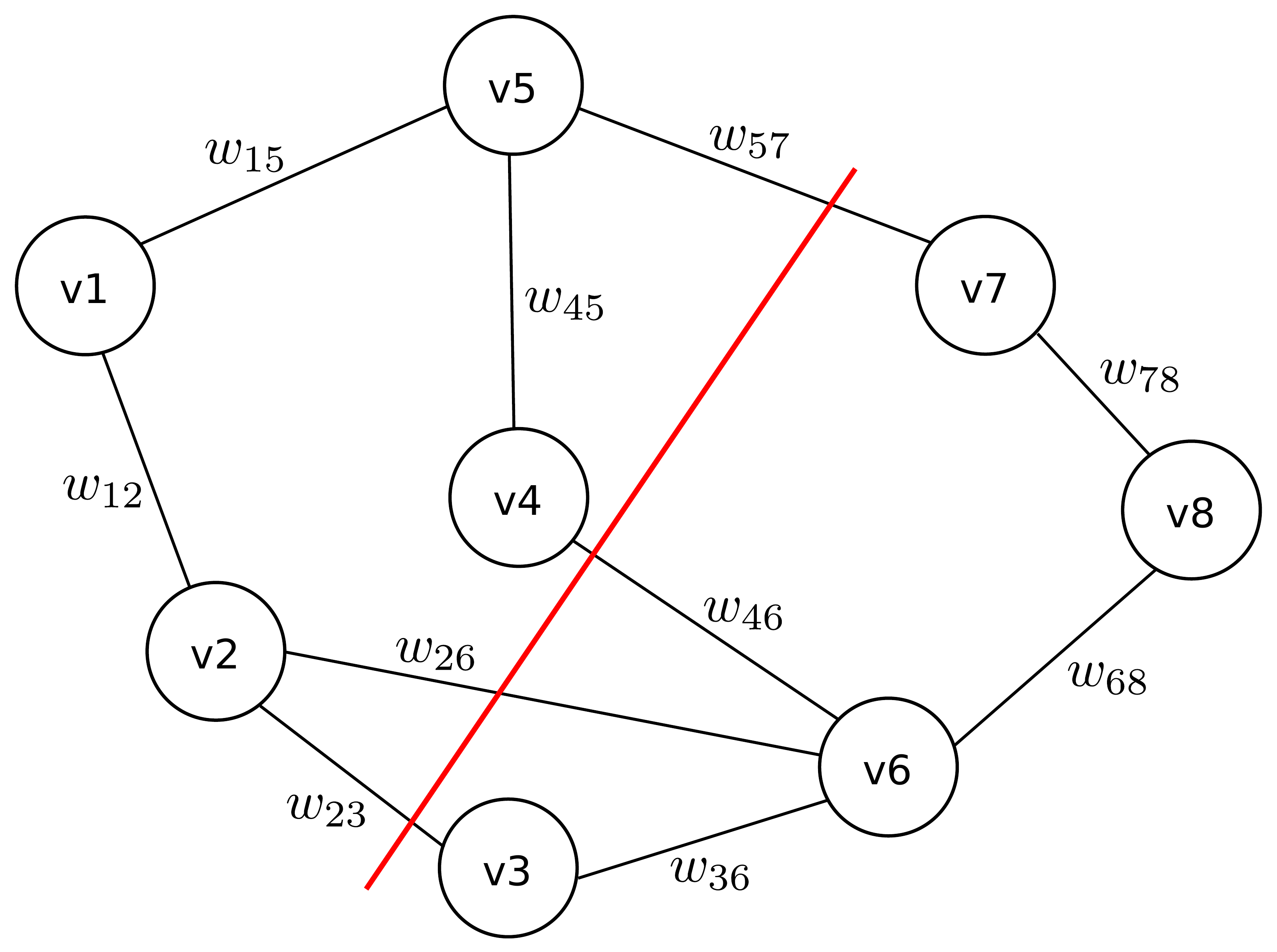}

\caption{Example of a Max-Cut problem. The red line indicates a proposed decomposition
(``cut'') into two graphs. The problem consists of bisecting the
given graph while maximizing the sum of weights along the cut.}
\end{figure}

It is perhaps surprising, given how ubiquIt\^{o}us systems that are
described by the Ising model are in nature, that it could theoretically
be used to facilitate solving all these computational problems. In
light of this, is it possible to build a (special-purpose) computational
machine, in which the calculation is ``carried out'' by magnetic
spins and their interactions with each other and with an external
magnetic field? 

For such a machine, one would have to be able to
\begin{itemize}
\item accurately set the interaction matrix $\mathbf{J}$ and external magnetic
field $h_{i}$ to arbitrary values
\item significantly reduce or mitigate the influence of external perturbations,
such as thermal fluctuations
\item accurately determine the spin states at the end of the ``computation
phase''
\end{itemize}
These requirements alone already pose significant challenges if one
attempted to use the magnetic spins of single atoms for which the
Ising model was original formulated. Additionally, it might be desirable
to control the initial state of the system as well as have some sort
of mechanism to increase the chance of the system evolving into its
ground state instead of a local minimum, which would constitute additional
challenges.

\subsection{CIM architectures}

In the setup of the Coherent Ising machine, a nonlinear material is
embedded in a ring cavity. Instead of using multiple DPOs to represent
the different spin states, the DPO is operated (pumped) in a pulsed
way such that all spin states of the system are represented by the
same DPO at different times. This means, for an Ising model with $N$
spins, the DPO will represent the spin states $\sigma_{1}$ during
the first pulse, $\sigma_{2}$ during the second pulse, etc., eventually
representing $\sigma_{N}$, before representing $\sigma_{1}$ again
in the next pulse.

The obvious missing ingredient is the interaction between spins, as
specified by the $\mathbf{J}$ matrix as well as with the external
magnetic field $h_{i}$. There are, as of the writing of this article,
two ways by which this is achieved.

The first one is through optical delay lines \citep{Marandi_CIM_Nature2014}.
Here, a part of the signal is extracted with an output coupler, amplified
by a phase-sensitive amplifier and led through a number of optical
lines before being fed back into the ring cavity with an injection
coupler. These optical lines are adjusted in length such that two
different optical pulses are produced when the delayed signal is fed back into
the cavity. Within the delay lines, the signal is adjusted through
amplification and phase shift to match the corresponding element of
the $\mathbf{J}$ matrix.

In the second scheme \citep{McMahon2016}, a part of the signal is
continuously measured via a homodyne detector. The appropriate feedback
signal is then calculated electronically. Based on this, the feedback
signal is generated separately through and intensity modulator and
a phase modulator acting on the laser beam which feeds the pump pulse.
The injected signal is fed back into the ring cavity. The calculation
of the feedback signal is carried out via a field-programmable gate array (FPGA)
to minimize computation
times. It might seem counterintuitive to use electronic circuitry,
here an FPGA, for the calculation of the feedback signal - after all,
isn't the goal of the Coherent Ising machine to design a computational
device based on physical processes other than (semiconductor) electronics?
However, the FPGA only computes a part of the problem, namely the
magnitude of the interaction strength, while the rest of the computation
still happens ``inside'' the ring cavity. On the other hand, the
measurement-feedback based CIM architecture has significant advantages
over the optical delay lines architecture as well. 

The main advantage of the optical delay lines architecture are very
fast operating times since no complicated logical gates, such as an
FPGA, are involved in the calculation of the feedback strength. However,
for an Ising model with $N$ spins and a dense $\mathbf{J}$ matrix,
up to $N-1$ optical delay lines are required. Hence, there are inherent
limitations on scalability. Conversely, for the measurement-feedback
architecture, the FPGA poses a bottleneck in operating time. At the
same time, the system can be scaled up to include a large number of
spin states very easily due to the absence of optical delay lines,
which was demonstrated recently by an experiment of a CIM involving
$100,000$ spin states \citep{honjo2021100}. Additionally, the measurement-feedback
scheme provides greater flexibility, which makes it possible to simulate
more exotic systems, such as Ising-like models that include interactions
between 3 or more spins, and sophisticated protocols to increase the
likelihood of reaching the ground state.

\begin{figure}

\includegraphics[width=0.95\columnwidth]{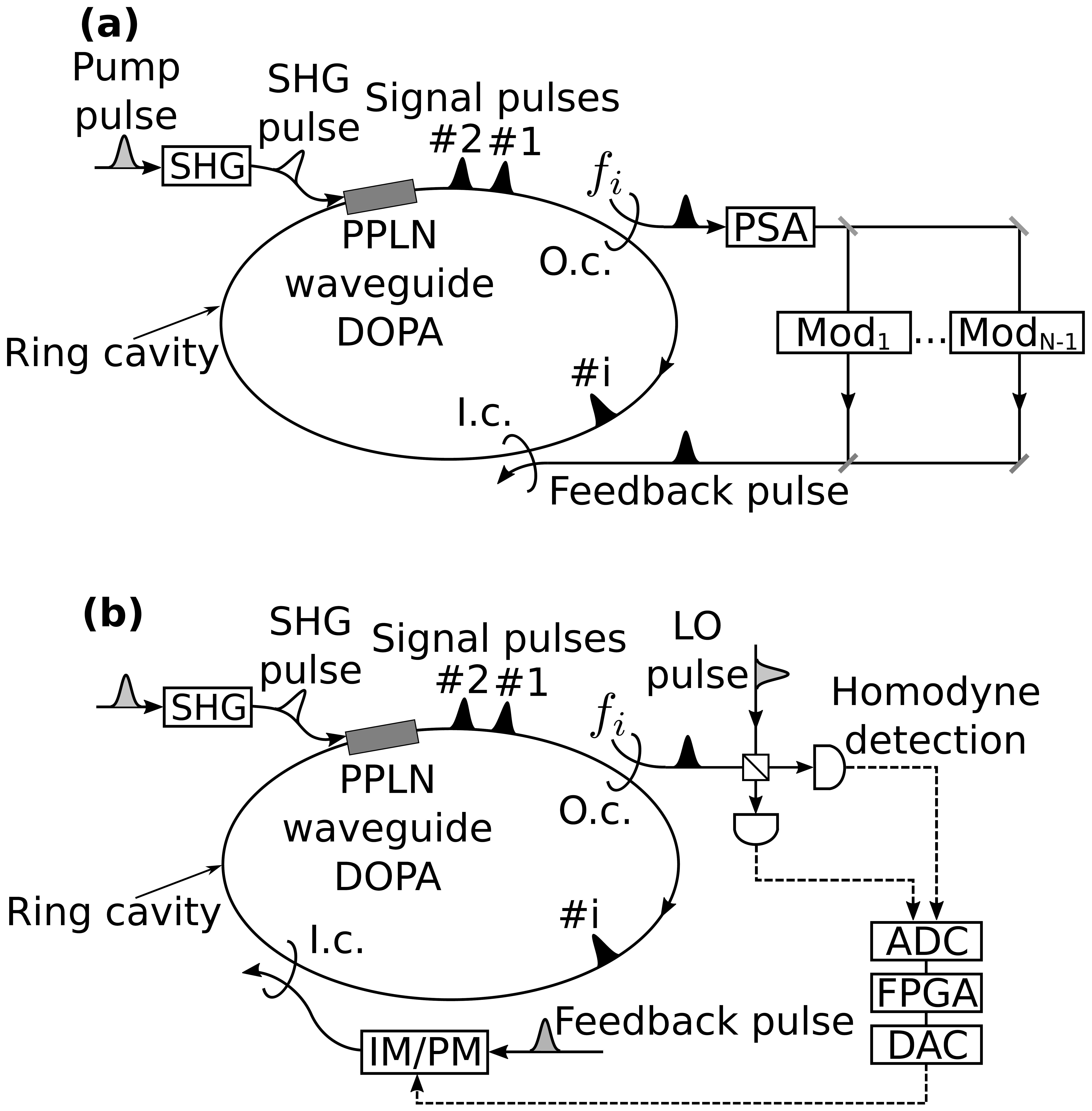}\caption{Two different CIM architectures. (a) shows the optical delay lines
type of CIM. Here, the feedback is generated by redirecting part of
the signal pulses through optical lines adjusted in length to match
up with different pulses at the injection coupler site. (b) shows
the measurement-feedback architecture, where the pulses are measured
by a homodyne detector. The measurement is digitized and the feedback
is calculated via an FPGA. Based on the calculation, an output pulse
is generated which is redirected into the ring cavity. Here, ``O.c.''
and ``I.c.'' stand for ``Output coupler'' and ``Injection coupler'',
respectively. Redrawn, following original published figure in \citep{yamamoto2017coherent}.}

\end{figure}

In this paper, we obtain a complete quantum description and a way
of simulating the measurement-feedback type of the CIM.

In the measurement-feedback architecture, the signal states are continuously
measured. Based on the measurement result, an FPGA calculates the
appropriate feedback based on the Ising model terms. Based on this
calculation, a separate signal is created and injected into the cavity.
Due to the continuous homodyne measurement that the signal states
are subject to, the wave-function experiences a continuous partial
state collapse conditional on the measurement outcome. This makes
the formulation of the system equations for the MFB-type CIM a more
complex task compared to the ODL architecture, as it involves the
theory of measurement-feedback quantum systems. In the following sections
we first describe a simple model of the individual DPO components,
and a general approach describing how to couple these via quantum
measurement-feedback theory.

\section{Degenerate parametric oscillator}

In place of physical spins, the Coherent Ising machine uses a degenerate
parametric oscillator (DPO)\citep{Drummond_OpActa1981,drummond2014quantum}.
In present experiments each DPO is a multi-mode, pulsed system due
to its traveling-wave nature \citep{raymer1991limits,werner1995ultrashort}.
However, to simplify the theory, it is common to use a single-mode
intra-cavity model. This treats each DPO as a single super-mode, which
is often valid classically \citet{hamerly2016reduced,roy2022temporal},
although a full multi-mode treatment is required to treat all quantum
noise effects, even in mode-locked systems \citet{drummond1997phase,patera2010quantum}.
While simpler than current fiber-optic experiments, the single-mode
model treats the most important features. It could in principle be
implemented more precisely in future experiments.

\subsection{Single-mode DPO theory}

As discussed above, we regard the CIM as a network of single-mode
DPOs. Each is essentially a $\chi^{\left(2\right)}$ nonlinear medium
embedded in an optical or microwave cavity. It is driven (pumped)
by a laser at frequency $\omega_{p}$. Due to the nonlinear medium,
parametric down-conversion can occur which leads to the creation of
two photons with frequencies $\omega_{s}$ and $\omega_{i}$. Subsequently,
we assume that $\omega_{i}=\omega_{s}=\omega_{p}/2$ and that the
cavity is resonant to both $\omega_{p}$ and $\omega_{s}$.

\begin{figure}
\includegraphics[width=0.9\columnwidth]{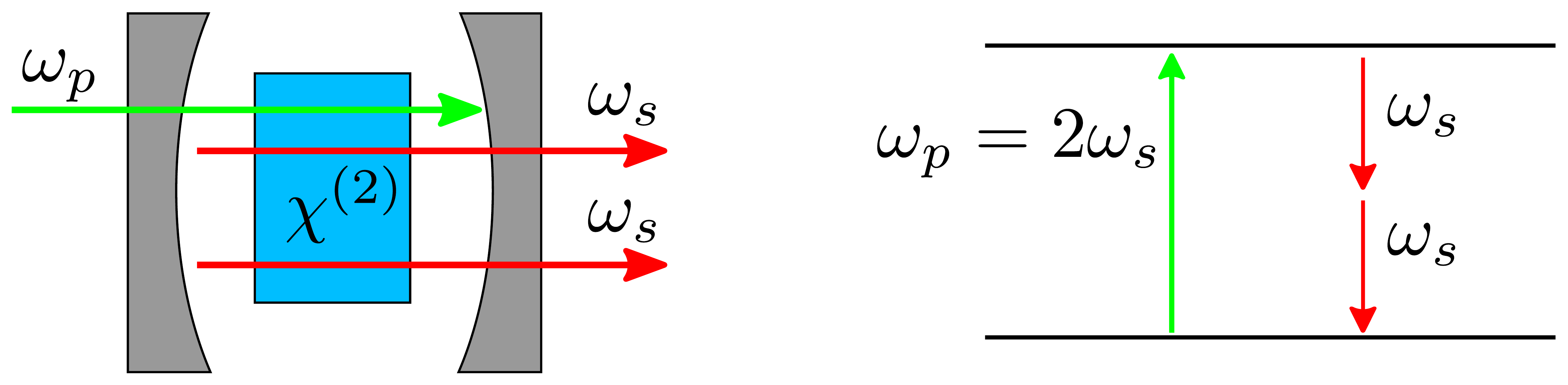}\caption{Schematic figure of a degenerate parametric oscillator on resonance.
Through parametric down-conversion taking place inside the optical
cavity, absorption of a single pump photon results in two signal photons
with half its frequency.}

\end{figure}

How can a nonlinear medium inside an optical cavity then take the
place of a discrete magnetic spin? 

In order to understand this, we first consider a DPO driven by a pump
field with induced amplitude $\mathcal{E}_{p}$, which is subject
to a decay rate $\gamma_{p}$, while photons created through parametric
down-conversion, which we subsequently call the signal field, are
subject to a decay rate $\gamma_{s}$. The DPO Hamiltonian is
\begin{eqnarray}
H_{DPO} & = & i\hbar\frac{\kappa}{2}\left[a_{p}\left(a_{s}^{\dagger}\right)^{2}-a_{p}^{\dagger}a_{s}^{2}\right]\,\nonumber \\
 &  & +i\hbar\left[\mathcal{E}_{p}a_{p}^{\dagger}-\mathcal{E}_{p}^{*}a_{p}\right]
\end{eqnarray}
where $a_{p}$, $a_{s}$ are the pump and signal field operators,
respectively and $\kappa$ is a nonlinearity parameter of the medium.
Before using a more complete quantum description later, we first consider
the evolution of the system in a classical picture. For this, we first
write down the Heisenberg-Langevin equations for the expectation values
of $a_{p}$, $a_{s}$, with the definitions that $\alpha=\left\langle a_{s}\right\rangle $
and $\alpha_{p}=\left\langle a_{p}\right\rangle $. 

To give an initial intuition about the behavior, we start by assuming
that expectation values factorize coherently, $\left\langle a_{s}^{\dagger}a_{p}\right\rangle =\alpha_{s}^{*}\alpha_{p}$,
and $\left\langle a_{s}^{2}\right\rangle =\alpha_{s}^{2}$ ,  which
gives
\begin{eqnarray}
\frac{d}{dt}\alpha_{s} & = & -\gamma_{s}\alpha_{s}+\kappa\alpha_{s}^{*}\alpha_{p}\nonumber \\
\frac{d}{dt}\alpha_{p} & = & \mathcal{E}_{p}-\gamma_{p}\alpha_{p}-\frac{1}{2}\kappa\alpha_{s}^{2}\,.\label{eq:DOPO_class}
\end{eqnarray}

One finds three steady-state solutions for Eq (\ref{eq:DOPO_class}):
\begin{eqnarray}
\alpha_{s} & = & 0\nonumber \\
\alpha_{p} & = & \mathcal{E}_{p}/\gamma_{p}\label{eq:DOPO_class_sol1}
\end{eqnarray}
as well as
\begin{eqnarray}
\alpha_{s} & = & \pm\sqrt{\frac{2}{\chi}\left[\mathcal{E}_{p}-\frac{\gamma_{p}\gamma_{s}}{\kappa}\right]}\nonumber \\
\alpha_{p} & = & \frac{\gamma_{s}}{\kappa}\,.\label{eq:DOPO_class_sol2}
\end{eqnarray}

A second-derivative test reveals that Eq (\ref{eq:DOPO_class_sol1})
is the only stable steady-state solution for $\mathcal{E}_{p}<\mathcal{E}_{p,th}$,
with $\mathcal{E}_{p,th}$ called the threshold pump strength defined
as $\mathcal{E}_{p,th}=\frac{\gamma_{s}\gamma_{p}}{\kappa}$, whereas
for $\mathcal{E}_{p}>\mathcal{E}_{p,th}$, Eq (\ref{eq:DOPO_class_sol1})
is an unstable solution and Eq (\ref{eq:DOPO_class_sol2}) are both
stable solutions. In the Coherent Ising machine, the two distinct
solutions of the DPO operated in the above-threshold regime take the
place of the discrete spin states.

\subsection{Quantum dynamics in phase-space}

The CIM is operated in a pulsed way with time-multiplexed spin states.
It nevertheless lends itself to a description of multiple DPO states
interacting simultaneously. This has an enormous Hilbert space. Conventional
number state expansions cannot be used in these cases, due to the
exponentially large basis set. It is therefore essential to use a
probabilistic approach in phase-space. This has been used in a number
of very large-scale quantum simulations \citep{drummond2016quantum,drummond2022simulating}.

Here, we want to analyze the phase-space dynamics of the system in
detail. Before looking at the multi-spin case, we will summarize known
results for the single-DPO system \citep{Drummond_OpActa1981}. We
consider the scenario where the DPO is driven by an induced pump rate
of $\mathcal{E}_{p}$ and the pump and signal field are subject to
a decay rate of $\gamma_{p}$ and $\gamma_{s}$, respectively. The
system evolution is described by the quantum master equation
\begin{eqnarray}
\frac{d}{dt}\rho & = & \gamma_{p}\mathcal{D}\left[a_{p}\right]\rho+\gamma_{s}\mathcal{D}\left[a_{s}\right]\rho+\frac{1}{i\hbar}\left[H_{DOPO},\rho\right]\,.\label{eq:DOPO_master}
\end{eqnarray}

The non-unitary evolution or mode damping is described by the super-operator
$\mathcal{D}\left[c\right]\rho\equiv2c\rho c^{\dagger}-\left(c^{\dagger}c\rho+\rho c^{\dagger}c\right)$,
which treats loss through the mirrors of the DPO model cavity, or
more general types of loss in the CIM experiment. The above equation
is now studied through its equivalent positive-P phase-space representation
\citep{Drummond_generalizedP1980}. This represents the density matrix
through an exact expansion in terms of general off-diagonal coherent-state
projectors,
\begin{equation}
\rho=\int P\left(\boldsymbol{\alpha},\boldsymbol{\beta}\right)\frac{\left|\bm{\alpha}\right\rangle \left\langle \bm{\beta}^{*}\right|}{\left\langle \bm{\beta}^{*}\right|\left.\bm{\alpha}\right\rangle }d^{2}\bm{\alpha}d^{2}\bm{\beta}\,.\label{eq:posP}
\end{equation}

The positive-P representation is chosen here over other representations
firstly because it is strictly non-negative, has a probabilistic interpretation,
and exists for all quantum states. It also results in a second-order
Fokker-Planck equation (FPE) with positive-definite diffusion that
has a corresponding stochastic process. This is achieved without the
necessity to remove (truncate) higher-order derivative terms, which
is important because equations with higher-order derivatives do not
have a stochastic equivalent. 

Mapping Eq (\ref{eq:DOPO_master}) to the positive-P representation
using standard operator identities \citep{Drummond_OpActa1981} yields
:
\begin{eqnarray}
\frac{dP}{dt} & = & \left\{ \gamma_{p}\left(\frac{\partial}{\partial\alpha_{p}}\alpha_{p}+\frac{\partial}{\partial\beta_{p}}\beta_{p}\right)+\gamma_{s}\left(\frac{\partial}{\partial\alpha_{s}}\alpha_{s}+\frac{\partial}{\partial\beta_{s}}\beta_{s}\right)\right.\nonumber \\
 &  & -\kappa\frac{\partial}{\partial\alpha_{s}}\left(\alpha_{s}\beta_{p}\right)-\kappa\frac{\partial}{\partial\beta_{s}}\left(\beta_{p}\alpha_{s}\right)\nonumber \\
 &  & +\frac{\partial}{\partial\alpha_{p}}\left(\frac{\kappa}{2}\alpha_{s}^{2}-\mathcal{E}_{p}\right)+\frac{\partial}{\partial\beta_{p}}\left(\frac{\kappa}{2}\beta_{s}^{2}-\mathcal{E}_{p}\right)\,\nonumber \\
 &  & \left.+\frac{\kappa}{2}\left[\frac{\partial^{2}}{\partial\alpha_{s}^{2}}\alpha_{p}+\frac{\partial^{2}}{\partial\beta_{s}^{2}}\beta_{p}\right]\right\} P\,,\label{eq:DOPO_FPE}
\end{eqnarray}
where $\bm{\alpha}\equiv\left(\alpha_{s},\alpha_{p}\right)$ and similarly
for $\bm{\beta}$ .

Although this equation has a diffusion term which is not positive-definite,
the non-orthogonal nature of the coherent--state expansion allows
one to obtain an equivalent, positive-definite FPE, which can then
be mapped into equivalent stochastic equations. Eq (\ref{eq:DOPO_FPE})
can be expressed through its corresponding set of stochastic differential
equations (SDEs), which are
\begin{eqnarray}
\frac{d}{dt}\alpha_{s} & = & \left(-\gamma_{s}\alpha_{s}+\kappa\alpha_{p}\beta_{s}\right)+\sqrt{\kappa\alpha_{p}}\xi_{1}\nonumber \\
\frac{d}{dt}\beta_{s} & = & \left(-\gamma_{s}\beta_{s}+\kappa\alpha_{s}\beta_{p}\right)+\sqrt{\kappa\beta_{p}}\xi_{2}\nonumber \\
\frac{d}{dt}\alpha_{p} & = & \mathcal{E}_{p}-\gamma_{p}\alpha_{p}-\frac{\kappa}{2}\alpha_{s}^{2}\nonumber \\
\frac{d}{dt}\beta_{p} & = & \mathcal{E}_{p}-\gamma_{p}\beta_{p}-\frac{\kappa}{2}\beta_{s}^{2}\,.
\end{eqnarray}

The equations at this stage can be interpreted as either Stratonovich
or It\^{o} SDEs, since there is no difference between the main two
types of stochastic calculus \citep{gardiner2004handbook} for these
equations. The terms $\xi_{1}$ and $\xi_{2}$ are delta-correlated
independent delta-correlated Gaussian noises, so that:
\begin{equation}
\left\langle \xi_{i}\left(t\right)\xi_{j}\left(t'\right)\right\rangle =\delta_{ij}\delta\left(t-t'\right).\label{eq:Gaussian_and_deltacorrelated}
\end{equation}

Usually, the pump field decay rate is much higher than the signal
field decay rate, which merits the adiabatic approximation that $\frac{d}{dt}\alpha_{p}=\frac{d}{dt}\beta_{p}=0$.
Assuming this, and defining $\chi\left(\alpha\right)=\frac{\kappa\mathcal{E}_{p}}{\gamma_{p}}-\frac{\kappa^{2}}{2\gamma_{p}}\alpha^{2}$,
the above SDEs reduce adiabatically \citep{gardiner1984adiabatic}
to a simpler It\^{o}-type SDE:
\begin{eqnarray}
\frac{d}{dt}\alpha_{s} & = & \left(-\gamma_{s}\alpha_{s}+\chi\left(\alpha_{s}\right)\beta_{s}\right)+\sqrt{\chi\left(\alpha_{s}\right)}\xi_{\alpha}\nonumber \\
\frac{d}{dt}\beta_{s} & = & \left(-\gamma_{s}\beta_{s}+\chi\left(\beta_{s}\right)\alpha_{s}\right)+\sqrt{\chi\left(\beta_{s}\right)}\xi_{\beta}\,.\label{eq:DOPO_SDE_be}
\end{eqnarray}

Based on this, one can reconstruct an adiabatic quantum master equation
and find that
\begin{eqnarray}
\frac{d\rho}{dt} & = & \frac{1}{i\hbar}\left[H_{s},\rho\right]+\left[\gamma_{s}\mathcal{D}\left[a_{s}\right]+\frac{\kappa^{2}}{4\gamma_{p}}\mathcal{D}\left[a_{s}^{2}\right]\right]\rho\label{eq:DOPO_adiab_master}
\end{eqnarray}
where $H_{s}\equiv i\hbar\frac{\kappa\epsilon_{p}}{2\gamma_{p}}\left[\left(a_{s}^{\dagger}\right)^{2}-a_{s}^{2}\right]\,.$
It is also possible to carry out the adiabatic elimination from the
master equation, giving an identical result \citep{carmichael2009statistical}.

The terms of Eqs. (\ref{eq:DOPO_SDE_be}) illustrate the different
physical processes happening simultaneously in the DPO cavity. The
first term corresponds to linear decay, the second term to linear
gain due to driving, and the third term to non-linear gain saturation.
The additional stochastic terms are due to quantum noise. However,
individual trajectories do not necessarily correspond to individual
experimental outcomes. From the expansion of Eq (\ref{eq:posP}),
we see that averages over many trajectories are required to reconstruct
even a single quantum state. Hence there is no general one-to-one
correspondence between trajectories and experimental outcomes. Yet,
due to the macroscopic nature of the ground state of the CIM, and
the microscopic coherent state variance, we conjecture that one can
regard the sign of the \emph{final} coherent state output at any one
site as giving the spin orientation at that site. This is confirmed,
at least for the current parameter values, by comparisons between
the conditional and unconditional results for our simulations.

\subsection{Relationship with neural networks}

Before analyzing the CIM in a comprehensive, fully quantum physical
description, we briefly want to point out its relation to neural networks.
To do this, we take Eqs. (\ref{eq:DOPO_SDE_be}) and make a number
of modifications. These simplify the equations to a classical model,
which takes us very far away from the world of quantum physics:
\begin{itemize}
\item We completely ignore the non-deterministic terms $\xi_{\alpha}$,
$\xi_{\beta}$. In other words, instead of a stochastic differential
equation, we are looking at a completely deterministic ordinary
differential equation.
\item We assume that $\alpha=\beta$, thus reducing the DPO to a single
equation of motion
\item We assume our phase-space variables are strictly real-valued
\end{itemize}
Further, we consider a set of $N$ DPOs, corresponding to $N$ spins
in an Ising model. Additionally, each DPO experiences an additional
term that drives the signal mode. This term corresponds to the signal
injected by the optical delay lines or the feedback signal in the
measurement-feedback architecture and is proportional to the interaction
term in the Ising model Hamiltonian, via a proportionality factor
$\zeta$. For simplicity, we assume the external magnetic field (Zeeman
term) to be zero.

Putting these assumptions together gives
\begin{eqnarray}
\frac{d\alpha_{i}}{dt} & = & \zeta\sum_{j}J_{ij}\alpha_{j}+\left(\frac{\kappa\mathcal{E}_{p}}{\gamma_{p}}-\gamma_{s}\right)\alpha_{i}-\frac{\kappa^{2}\alpha_{i}^{3}}{2\gamma_{p}}\,.\label{eq:CIM_class}
\end{eqnarray}

Let us compare the above equation with the concept of a neural network.
A neural network broadly speaking consists of the following ingredients\citep{Yamamoto2020,Stern2014}:
\begin{itemize}
\item A number of units, sometimes referred to as ``cells'', carrying
a real value. These units can be arranged in multiple layers, in which
case the network is referred to as a deep neural network, though other
arrangements are possible as well.
\item A connection between the units across different layers or units in
general
\item A non-linear output function associated with each unit. Though this
might seem like an optional detail, it is actually a crucial component.
Without nonlinearities, the network would be reduced to a linear function
of its input values, regardless of its number of layers and internal
complexity.
\end{itemize}
We can recognize all of these ingredients in Eq (\ref{eq:CIM_class}).
The variables $\alpha_{i}$ take the place of the network units. The
$\sum_{j}J_{ij}\alpha_{j}$ term represents the connection between
units while the remaining terms represent a third-order nonlinearity.
Since there are no layers here and the units are connected to each
other, Eq (\ref{eq:CIM_class}) is most akin to a recurrent neural
network (RNN) architecture.

Eq (\ref{eq:CIM_class}), though it is an incomplete description of
the system since it does not take into account the quantum nature
of the Coherent Ising machine, it can nevertheless be a useful tool
to analyze the convergence properties that can be expected.

Eq (\ref{eq:CIM_class}) follows the potential function
\begin{eqnarray}
\phi\left(\alpha_{i}\right) & = & -\zeta\sum_{i,j}J_{ij}\alpha_{i}\alpha_{j}-\frac{1}{2}\left(\frac{\kappa\mathcal{E}_{p}}{\gamma_{p}}-\gamma_{s}\right)\sum_{i}\alpha_{i}^{2}\nonumber \\
 &  & +\frac{\kappa^{2}}{8\gamma_{p}}\sum_{i}\alpha_{i}^{4}\,.\label{eq:CIM_class_pot}
\end{eqnarray}

The exact shape of the energy landscape strongly depends on $\mathbf{J}$.
The configuration that classically solves the Ising problem constitutes
the energetic minimum, however different configurations can (and typically
do) manifest as local minima, which poses obstacles in determining
the ground state. Stochasticity terms originating from quantum effects
and from environment interactions will contribute to exploring the
energy landscape. However, like a true Ising model, it is possible
and quite probable for the system to evolve into a local minimum,
which is not necessarily the global optimal minimum. The likelihood
of this is not only determined by the Ising problem itself, but also
by the minimization strategy that is employed.

A simple minimization strategy consists in linearly ramping up the
pump rate $\mathcal{E}_{p}$. Similar to the single DPO, which experiences
a bifurcation when $\mathcal{E}_{p}$ reaches $\mathcal{E}_{p,th}$,
minima in the energy landscape in Eq (\ref{eq:CIM_class_pot}) will
appear or become more pronounced with higher $\mathcal{E}_{p}$. In
comparison to a single DPO, the pump strength at which local minima
appear is usually lower for a network of coupled DPOs. For a more
gradual increase in $\mathcal{E}_{p}$, the local minima appear more
slowly which makes the minimization a more adiabatic one, which in
turn increases the likelihood of finding the global minimum for the
price of an overall longer simulation time. Once a certain pump strength
is reached, the system experiences a so-called ``freeze-out'', from
which the spin states do not change any more. This effectively marks
the end of the simulation process.

Besides this simple strategy, more sophisticated ones exist as well,
which can result in a higher likelihood of reaching the ground state. 

\section{Quantum Feedback CIM}

\subsection{Measurement-feedback theory\label{subsec:Measurement-feedback-theory}}

The framework of measurement-feedback systems was developed in the
early 1990s \citep{wiseman1994quantum,wiseman1993quantum,Diosi_PRA1989,carmichael1993quantum,dalibard1992wave,molmer1993monte}
and is based on earlier theories of measurement \citep{kraus1971general}.
We will review the basic concepts first before applying them to the
CIM.

We start by considering a quantum system described by the density
matrix $\rho$ subject to the evolution
\begin{eqnarray}
\frac{d}{dt}\rho\left(t\right) & = & \mathcal{L}\rho\left(t\right)\,,\label{eq:master_Lonly}
\end{eqnarray}
where $\mathcal{L}$ includes both unitary and non-unitary terms.

\subsection{Quantum stochastic measurement equations}

Next, suppose that a continuous quantum measurement of an Hermitian
operator $c+c^{\dagger}$ is carried out, which in this case is the
output field proportional to the quadrature $\hat{X}=a+a^{\dagger}$.
The measurement outcome at time $t$ is determined not only by the
state $\rho\left(t\right)$, but also of the quantum noise that is
 introduced in the measurement process and cannot be eliminated by
an improved measurement apparatus. The system state $\rho\left(t\right)$
partially collapses based on the measurement outcome. Because the
measurement outcome is not predictable, due to quantum noise, the
state of $\rho$ at time $t+dt$ is not predictable either. 

The measurement outcome and the evolution of $\rho$ become intertwined.
To make this problem more tractable, one introduces the \textit{conditional}
state $\rho_{c}\left(t\right)$ and measurement outcome $I_{c}\left(t\right)$.
Here, conditional refers to a specific \textit{realization} of the
quantum measurement noise. As well as generating a directly measured
outcome, the measurement changes the quantum state, in a process described
by the generalized theory of measurement effects and operations \citep{kraus1971general,caves1994quantum}.

Assuming a detector with perfect efficiency is used, these are obtained
in this case \citep{wiseman1993quantum} as:
\begin{eqnarray}
\frac{d}{dt}\rho_{c}\left(t\right) & = & \left[\mathcal{L}+\xi\left(t\right)\mathcal{H}\left[c\right]\right]\rho_{c}\left(t\right)\,.\label{eq:CIM_H_nofb}
\end{eqnarray}
Here $\xi\left(t\right)$ is the fluctuating part of the feedback
current:
\begin{equation}
I_{c}\left(t\right)=\left\langle c+c^{\dagger}\right\rangle _{c}\left(t\right)+\xi\left(t\right),\label{eq:Ic_t}
\end{equation}
while $\left\langle c+c^{\dagger}\right\rangle _{c}\left(t\right)\equiv\text{Tr}\left[\left(c+c^{\dagger}\right)\rho_{c}\left(t\right)\right]$,
and $\xi\left(t\right)$ is a Gaussian delta-correlated noise as in
Eq (\ref{eq:Gaussian_and_deltacorrelated}). The super-operator $\mathcal{H}$$\left[\cdot\right]$,
which describes the effects of the measurement and ensures preservation
of the trace of $\rho_{c}$, is called the innovation operator, and
is defined as
\begin{eqnarray}
\mathcal{H}\left[c\right]\rho & \equiv & c\rho+\rho c^{\dagger}-\text{Tr}\left[c\rho+\rho c^{\dagger}\right]\rho\,.
\end{eqnarray}

The term $\xi\left(t\right)$ represents the quantum measurement noise.
Eq (\ref{eq:CIM_H_nofb}) is to be understood as an It\^{o}-type
stochastic differential equation (SDE), but with operator rather then
c-number stochastic variables. \textcolor{green}{}

\subsection{It\^{o} and Stratonovich equations}

For use in simulating conditional feedback, we will use the equivalent
Stratonovich master equation. These have the advantage that they satisfy
the standard rules of calculus, and are generally simpler to integrate.
The theory of this equivalence is well understood \citep{gardiner2004handbook}.
A generic multivariate, $m$-dimensional Markovian stochastic process
has an It\^{o}-type equation
\begin{eqnarray}
\frac{d}{dt}\mathbf{X}^{\left(I\right)} & = & \mathbf{A}\left(\mathbf{X}\right)+\mathbf{B}\left(\mathbf{X}\right)\mathbf{\bm{\xi}}\left(t\right)\,,
\end{eqnarray}
 where $\mathbf{A}$ is an $m$-dimensional vector, $\mathbf{\mathbf{B}}$
is an $m\times n$ dimensional matrix, and $\mathbf{\bm{\xi}}$ is
an $n$-dimensional noise correlated according to Eq (\ref{eq:Gaussian_and_deltacorrelated}).
The corresponding Stratonovich-type stochastic differential equation
can be found via
\begin{eqnarray}
\frac{d}{dt}\mathbf{X}^{\left(S\right)} & = & \mathbf{A}\left(\mathbf{X}\right)+\mathbf{C}\left(\mathbf{X}\right)+\mathbf{B}\left(\mathbf{X}\right)\mathbf{\bm{\xi}}\left(t\right)\,.\nonumber \\
\label{eq:Strat_corr_c_numbers}
\end{eqnarray}

Here, $\mathbf{C}\left(\mathbf{X}\right)$ is called the Stratonovich
correction term, where:
\begin{equation}
C_{i}\left(\mathbf{X}\right)=-\frac{1}{2}\sum_{k,j}\frac{\partial B_{ik}}{\partial X_{j}}\left(\mathbf{X}\right)B_{jk}\left(\mathbf{X}\right)\,.
\end{equation}
For complex stochastic vectors, one can generalize this by expanding
in real and imaginary parts or by using Wirtinger calculus. Although
Eq (\ref{eq:CIM_H_nofb}) is an operator equation, the above transformation
rule can be applied by considering the quantum operators, including
the density operator, as large matrices. This way, one finds a Stratonovich
correction term as in Eq (\ref{eq:Strat_corr_c_numbers}). 

Hence, for complex stochastic matrices $X_{ij}$, if $\mu=\left(i,j\right)$
and $\nu=\left(i',j'\right)$, one obtains
\begin{equation}
C_{\mu}=-\frac{1}{2}\sum_{k,\nu}\left[B_{\nu k}\frac{\partial}{\partial X_{\nu}}+B_{\nu k}^{*}\frac{\partial}{\partial X_{\nu}^{*}}\right]B_{\mu k}\,.
\end{equation}

In the cases treated here, since $\mathcal{\mathcal{H}}\left[c\right]\rho_{c}$
is analytic in $\rho$, there is no extra Wirtinger term from the
conjugate derivative. The resulting Stratonovich correction corresponding
to the single-mode measurement operator $\mathcal{H}\left[c\right]\rho$
is:
\begin{equation}
C^{\mathcal{H}}\left[c\right]\rho_{c}=\left\langle c+c^{\dagger}\right\rangle _{c}\mathcal{H}\left[c\right]\rho_{c}-\frac{1}{2}\mathcal{H}\left[c^{2}\right]\rho_{c}+\left\langle c^{\dagger}c\right\rangle \rho_{c}-c\rho_{c}c^{\dagger}\,.\label{eq:C_H_term}
\end{equation}

\textcolor{green}{}Although reported previously \citet{hush2013controlling},
this result is not well-known, and we give the complete proof in the
Appendix. Hence, the equivalent Stratonovich-type master equation
is therefore obtained as:
\begin{eqnarray}
\frac{d}{dt}\rho_{c} & = & \left[\mathcal{L}+\xi\left(t\right)\mathcal{H}\left[c\right]+C^{\mathcal{H}}\right]\rho_{c}\left(t\right)\,.\label{eq:Strat_master_eq}
\end{eqnarray}
For our purposes, the operator $c$ is proportional to the operator
$a$ of the DPO signal state, which corresponds to homodyne detection.
It is important to note that since a measurement of $c$ is taking
place, there needs to be a loss term $\mathcal{D}\left[c\right]$
included in the operator $\mathcal{L}$. For example, if a homodyne
detection is being carried out, the operator $\mathcal{D}\left[a\right]$
is required to account for the fact that part of the signal leaves
the cavity for the detector. 

Realistically, the detector will have limited detection efficiency.
This can be accounted for by ``splitting up'' the fraction of the
signal which enters the detector into a fraction which decays without
being detected and a fraction which decays while being detected. This
will be demonstrated shortly when the measurement-feedback scheme
is applied to the CIM.

\subsection{Feedback master equations\label{subsec:Single_mode_feedback}}

We now wish extend the system so that it includes a feedback which
is applied to $\rho_{c}$ based on the measurement $I_{c}\left(t\right)$.
The feedback is expressed by a super-operator $\mathcal{K}$ which
depends on the feedback mechanism. This may be defined as a unitary
operator via $\mathcal{K}\rho\equiv\left[K,\rho\right]$, where $K$
is an arbitrary operator. We further limit ourselves to the case where
the feedback super-operator is a linear function of the measurement
result $I_{c}\left(t\right)$.

Between the measurement of the system state and the application of
feedback, there is always some delay time $\tau$. For example, in
the case of our MFB CIM, there is the propagation time between the
photodetector, the FPGA and the signal generator as well as the calculation
time of the FPGA. Hence, in a precise description, the feedback operator
would be proportional to a retarded measurement operator, i.e. $I_{c}\left(t-\tau\right)\mathcal{K}\rho\left(t\right)$.
To simplify things, we want to take the limit of $\tau\rightarrow0$. 

A naive approach to incorporate the feedback would then be simply
to add a feedback term $I_{c}\left(t\right)\mathcal{K}\rho_{c}\left(t\right)$
to Eq (\ref{eq:CIM_H_nofb}) to get
\begin{eqnarray}
\frac{d}{dt}\rho_{c}\left(t\right) & = & \left[\mathcal{L}+\xi\left(t\right)\mathcal{H}\left[c\right]+I_{c}\left(t\right)\mathcal{K}\right]\rho_{c}\left(t\right)\,.\label{eq:CIM_fb_false}
\end{eqnarray}

However, this approach poses a conceptual problem, and this equation
will not be used. Even in the limit $\tau\rightarrow0$, any feedback
based on a specific measurement happens \emph{after} the state collapse
of the wave-function in relation to this measurement outcome. When
formulating Eq (\ref{eq:CIM_fb_false}), we have not done anything
to take this causal delay into account. 

In fact, with the definition of $I_{c}\left(t\right)$, Eq (\ref{eq:CIM_fb_false})
indicates that the wave-function collapse happens simultaneously with
the feedback. In the following sub-sections, we describe the theory
that solves this problem.

The correct expression for the feedback equation can be found following
a derivation in \citet{wiseman1994quantum}. Here, the action of the
feedback is accounted for via an exponential term that acts from the
left on the remaining terms. This ensures the correct operator ordering
between $\mathcal{K}$ and $\mathcal{H}$, consistent with the fact
that feedback necessarily follows after the wave-function collapse
or measurement process.

We describe the general approach here, and use it to obtain a conditional
master equation for one mode, which will be generalized in the next
subsection. We use the definition of $I_{c}\left(t\right)$ and express
the feedback as well as Eq (\ref{eq:master_Lonly}) in differential
form to get:
\begin{eqnarray}
\rho_{c}\left(t+dt\right) & = & \exp\left[\left\langle c+c^{\dagger}\right\rangle _{c}\left(t\right)\mathcal{K}\cdot dt+\mathcal{K}\cdot dW\right]\cdot\nonumber \\
 &  & \left[1+\left(\mathcal{L}\cdot dt+dW\cdot\mathcal{H}\left[c\right]\right)\right]\rho_{c}\left(t\right)\,,\label{eq:master_fb_exp}
\end{eqnarray}
where $dW$ is the noise increment of $\xi\left(t\right)$ in time
$dt$, that is, $\xi\left(t\right)=dW/dt$.

One now expands the exponential in Eq (\ref{eq:master_fb_exp}) to
second order, expands the product and disregards terms of order $\mathcal{O}\left(dt^{3/2}\right)$
and above. We use the prescription that $dW^{2}\sim dt$. The approach
given here omits details that are given in the original literature
\citep{wiseman1993quantum,diosi1994comment}. This leads to an It\^{o}
equation,
\begin{eqnarray}
\frac{d}{dt}\rho_{c}\left(t\right) & = & \frac{1}{dt}\left[\rho_{c}\left(t+dt\right)-\rho_{c}\left(t\right)\right]\nonumber \\
 & = & \mathcal{L}\rho_{c}\left(t\right)+\mathcal{K}\left(c\rho_{c}\left(t\right)+\rho_{c}\left(t\right)c^{\dagger}\right)\nonumber \\
 &  & +\left\{ \frac{1}{2}\mathcal{K}^{2}+\xi\left(t\right)\left[\mathcal{H}\left[c\right]+\mathcal{K}\right]\right\} \rho_{c}\left(t\right)\label{eq:master_fb_true}
\end{eqnarray}

Like Eq (\ref{eq:CIM_H_nofb}), Eq (\ref{eq:master_fb_true}) is an
It\^{o}-type stochastic differential equation. As an alternative
approach of obtaining Eq (\ref{eq:master_fb_true}), Eq (\ref{eq:CIM_H_nofb})
can be transformed into its corresponding Stratonovich form. Following
the procedure described in the Appendix, which takes account of all
the Stratonovich corrections, one obtains under certain restrictions
for the form of the super-operators $\mathcal{H}$ and $\mathcal{K}$,
that:
\begin{eqnarray}
\frac{d}{dt}\rho_{c}^{(S)} & = & \left[\mathcal{L}+\left\langle c^{\dagger}c\right\rangle -\frac{1}{2}\mathcal{H}\left[cc\right]+I_{c}\left(\mathcal{H}\left[c\right]+\mathcal{K}\right)\right]\rho_{c}-c\rho_{c}c^{\dagger}\,,\nonumber \\
\label{eq:Stratonovich-single-mode-ME}
\end{eqnarray}
where $I_{c}\left(t\right)$ is given by Eq (\ref{eq:Ic_t}). 

\textcolor{blue}{}

Eqs. (\ref{eq:master_fb_true}) and (\ref{eq:Stratonovich-single-mode-ME})
describe the evolution of the system for a given measurement noise
outcome $\xi\left(t\right)$. However, in many cases we are not interested
in what happens for a specific noise realization, but rather what
happens over a range of many noise realizations. In such cases we
can average the It\^{o} master equation over the infinitely many
outcomes for $\xi\left(t\right)$ to obtain $\left\langle \frac{d}{dt}\rho_{c}\left(t\right)\right\rangle _{\xi\left(t\right)}=\frac{d}{dt}\left\langle \rho_{c}\left(t\right)\right\rangle _{\xi\left(t\right)}$. 

Defining $\rho\equiv\left\langle \rho_{c}\left(t\right)\right\rangle _{\xi\left(t\right)}$,
the average over this type of It\^{o} stochastic equation has the
effect of simply removing the noise terms, due to linearity and the
non-anticipating nature of It\^{o} calculus \citep{gardiner2004handbook}.
Therefore, this yields a much simpler equation:
\begin{eqnarray}
\frac{d}{dt}\rho & = & \mathcal{L}\rho+\mathcal{K}\left(c\rho+\rho c^{\dagger}\right)+\frac{1}{2}\mathcal{K}^{2}\rho\,.
\end{eqnarray}

\subsection{Master equation for multiple nodes}

We now consider a CIM with $N$ DPOs (spin states). We assume the
adiabatic approximation and label the spin states $a_{1},...,a_{N}$,
dropping the index $s$. The RHS of Eq (\ref{eq:DOPO_adiab_master}),
applied to all modes $a_{i}$, is equivalent to $\mathcal{L}\rho\left(t\right)$
in the framework outlined above. We now introduce a second channel
through which the signal decays with rate $\gamma_{m}$. The fraction
of the signal that decays through this channel shall be observed by
a homodyne detector with perfect efficiency. This way, a homodyne
detector with limited efficiency can be described by declaring that
the fraction not picked up by the detector decays through the channel
already present in Eq (\ref{eq:DOPO_adiab_master}). Thus, we have
to add another set of diffusive terms $\gamma_{m}\mathcal{D}\left[a_{i}\right]$. 

With $\gamma\equiv\gamma_{s}+\gamma_{m}$, these can be combined to
yield $\gamma\mathcal{D}\left[a_{i}\right]$. The operator subject
to the innovation operator $\mathcal{H}\left[\cdot\right]$ is $\sqrt{2\gamma_{m}}a_{i}$.
Thus, \emph{without} the feedback, we obtain the total master equation
\begin{eqnarray}
\frac{d}{dt}\rho & = & \frac{1}{i\hbar}\left[H_{s},\rho\right]+\sum_{i}\left\{ \gamma\mathcal{D}\left[a_{i}\right]\rho+\frac{\kappa^{2}}{4\gamma_{p}}\mathcal{D}\left[a_{i}^{2}\right]\rho\right\} \rho\nonumber \\
 &  & +\sum_{i}\mathcal{H}\left[\sqrt{2\gamma}a_{i}\right]\rho\nonumber \\
 & \equiv & \mathcal{L}\rho+\sum_{i}\mathcal{H}\left[\sqrt{2\gamma}a_{i}\right]\rho\,,
\end{eqnarray}
where 
\begin{equation}
H_{s}=i\hbar\frac{\kappa\epsilon_{p}}{2\gamma_{p}}\sum_{i}\left[\left(a_{i}^{\dagger}\right)^{2}-a_{i}^{2}\right]\,.
\end{equation}

This treats many modes in parallel, with measurement as well, but
there is no feedback included at this stage.

We now extend the framework outlined so far to include several modes
\emph{including} feedback. The interaction matrix $\mathbf{J}$ is
typically denser than a permutation matrix, in other words, a measurement
outcome for one mode $a_{i}$ will generally produce feedback in several
modes among $a_{1},...,a_{N}$ proportional to $J_{i1},...,J_{iN}$.
As a result, Eq (\ref{eq:master_fb_exp}) becomes
\begin{eqnarray}
\rho_{c}\left(t+dt\right) & = & \exp\left[\sum_{i}\mathcal{K}_{i}\sum_{j}J_{ij}\left[\left\langle c_{j}+c_{j}^{\dagger}\right\rangle _{c}dt+dW_{j}\right]\right]\times\nonumber \\
 &  & \left[\rho_{c}+\left(\mathcal{L}\cdot dt+\sum_{j}dW_{j}\cdot\mathcal{H}\left[c_{j}\right]\right)\rho_{c}\right],\label{eq:CIM_master_exp}
\end{eqnarray}
where time-dependent functions on the RHS are evaluated at $\left(t\right)$,
and $\mathcal{K}_{i}$ is the super-operator that generates feedback
for the $i$'th mode.

After expanding Eq (\ref{eq:CIM_master_exp}) and retaining all terms
of order $\mathcal{O}\left(dt\right)$, $\mathcal{O}\left(dW\right)$
and $\mathcal{O}\left(1\right)$, one finds
\begin{eqnarray}
\frac{d}{dt}\rho_{c} & = & \mathcal{L}\rho_{c}+\sum_{ij}\mathcal{K}_{i}\left(J_{ij}\left(c_{j}\rho_{c}+\rho_{c}c_{j}^{\dagger}\right)\right)\nonumber \\
 &  & +\frac{1}{2}\sum_{i,j,k}J_{ij}J_{ik}\mathcal{K}_{i}\mathcal{K}_{k}\rho_{c}\nonumber \\
 &  & +\sum_{i}\xi_{i}\left(t\right)\left[\mathcal{H}\left[c_{i}\right]+\sum_{j}J_{ij}\mathcal{K}_{j}\right]\rho_{c}\,.\label{eq:fb_total_w_noise}
\end{eqnarray}

This is an It\^{o} conditional master equation, which needs to be
solved relative to every noise realization. Consider a single entry
$J_{ij}$ from the interaction matrix $\mathbf{J}$. We want the feedback
to induce the signal $\zeta J_{ij}\left(a_{j}\rho+\rho a_{j}^{\dagger}\right)$
into the $i$-th mode.\textcolor{green}{{} }With the definition of $c_{j}=\sqrt{2\gamma_{m}}a_{j}$,
we find that
\begin{eqnarray}
\mathcal{\mathcal{K}}_{i}\rho & = & \frac{\zeta}{\sqrt{2\gamma_{m}}}\left[a_{i}^{\dagger}-a_{i},\rho\right]\,.
\end{eqnarray}

Using the techniques from Sec \ref{subsec:Single_mode_feedback},
the Stratonovich form of Eq (\ref{eq:fb_total_w_noise}) is found
to be
\begin{eqnarray}
\frac{d}{dt}\rho_{c}^{\left(S\right)} & = & \mathcal{L}\rho_{c}^{\left(S\right)}+\sum_{ij}\mathcal{K}_{i}J_{ij}\left\langle c_{j}+c_{j}^{\dagger}\right\rangle \nonumber \\
 &  & +\sum_{i}\xi_{i}\left(t\right)\left[\mathcal{H}\left[c_{i}\right]+\sum_{j}J_{ij}\mathcal{K}_{j}\right]\rho_{c}^{\left(S\right)}\nonumber \\
 &  & +\sum_{i}C^{\mathcal{H}}\left[c_{i}\right]\rho_{c}^{\left(S\right)}\label{eq:fb_total_w_noise_Strat}
\end{eqnarray}
with $C^{\mathcal{H}}$ given in Eq (\ref{eq:C_H_term}). 

For the It\^{o} form, given in Eq (\ref{eq:fb_total_w_noise}), averaging
over the noise outcomes simply removes the last term in Eq (\ref{eq:fb_total_w_noise}),
and yields
\begin{eqnarray}
\frac{d}{dt}\rho & = & \mathcal{L}\rho+\sum_{i}\mathcal{K}_{i}\sum_{j}J_{ij}\left(c_{j}\rho+\rho c_{j}^{\dagger}\right)\left(t\right)\nonumber \\
 &  & +\frac{1}{2}\sum_{i,j,k}J_{ij}J_{ik}\mathcal{K}_{i}\mathcal{K}_{k}\rho\,.
\end{eqnarray}

Finally, we have all ingredients for a full description of the MFB-type
CIM. Substituting all definitions, we get the total quantum master
equation
\begin{eqnarray}
\frac{d\rho}{dt} & = & \sum_{i}\left(\frac{\kappa\epsilon_{pi}}{2\gamma_{p}}\left[\left(a_{i}^{\dagger}\right)^{2}-a_{i}^{2},\rho\right]+\gamma\mathcal{D}\left[a_{i}\right]\rho\right.\nonumber \\
 &  & +\frac{\kappa^{2}}{4\gamma_{p}}\mathcal{D}\left[a_{i}^{2}\right]\rho+\zeta\sum_{j}J_{ij}\left[a_{i}^{\dagger}-a_{i},a_{j}\rho+\rho a_{j}^{\dagger}\right]\nonumber \\
 &  & \left.+\frac{\zeta^{2}}{4\gamma_{m}}\sum_{j,k}J_{ij}J_{ik}\left[a_{i}^{\dagger}-a_{i},\left[a_{k}^{\dagger}-a_{k},\rho\right]\right]\right).\label{eq:CIM_fb_master_full}
\end{eqnarray}

Such total master equations have been numerically solved then compared
to experiment in much simpler cases of laser cooling through feedback
\citep{bushev2006feedback}. In these studies, comparison to the full
conditional master equation was generally not carried out, due to
the computational and experimental complexity of recording and storing
the full measurement history for each feedback realization.

\subsection{Total phase-space simulations}

Even though very much simpler than the conditional master equation,
the total master equation, Eq (\ref{eq:CIM_fb_master_full}) is still
insoluble analytically, as far as we know. Treating it with orthogonal
state expansions is exponentially hard with large numbers of modes,
as in the CIM. It has only been carried out for small Hilbert spaces,
usually involving state truncation as well \citep{bushev2006feedback}. 

Despite this, it can be simulated via phase-space methods, using the
positive-P representation. This solves the exponential hardness problem
through probabilistic sampling. These techniques are known to be successful
in a number of similar cases with large bosonic Hilbert spaces \citep{drummond2016quantum}.
There is a known limitation, however. For low losses, high nonlinearities
and long time-evolution, boundary term errors can break the stochastic
equivalence \citep{Gilchrist_PRA1997,Smith1989}. While this can be treated
using stochastic gauge methods \citep{Deuar_RPA2002}, this is not
required for typical CIM parameters.

Using the standard rules, Eq (\ref{eq:CIM_fb_master_full}) translates
to the Fokker-Planck equation
\begin{eqnarray}
\frac{dP}{dt} & = & \left\{ \sum_{i}\left[\partial_{\alpha_{i}}\left(\gamma\alpha_{i}-\chi\left(\alpha_{i}\right)\beta_{i}\right)+\partial_{\alpha_{i}}^{2}\chi\left(\alpha_{i}\right)\right]\right.\nonumber \\
 &  & +\sum_{i}\left[\partial_{\beta_{i}}\left(\gamma\beta_{i}-\chi\left(\beta_{i}\right)\alpha_{i}\right)+\partial_{\beta}^{2}\chi\left(\beta_{i}\right)\right]\nonumber \\
 &  & +f^{2}\sum_{i,j,k}J_{ij}J_{ik}\left(\partial_{\alpha_{i}}+\partial_{\beta_{i}}\right)\left(\partial_{\alpha_{k}}+\partial_{\beta_{k}}\right)\nonumber \\
 &  & \left.-\sum_{i,j}\left(\partial_{\alpha_{i}}+\partial_{\beta_{i}}\right)\zeta J_{ij}\left(\alpha_{j}+\beta_{j}\right)\right\} P\,,
\end{eqnarray}
where $f=\zeta/\sqrt{2\gamma_{m}}$ , and $P\equiv P\left(\alpha_{1},\beta_{1},...,\alpha_{N},\beta_{N}\right)$. 

Translating this to a set of stochastic differential equations yields:
\begin{eqnarray}
\dot{\alpha}_{i} & = & \left[\epsilon_{i}-\gamma\alpha_{i}+\beta_{i}\chi\left(\alpha_{i}\right)\right]+\sqrt{\chi\left(\alpha_{i}\right)}\xi_{i}^{\alpha}+f\sum_{j}J_{ij}\xi_{j}\nonumber \\
\dot{\beta}_{i} & = & \left[\epsilon_{i}-\gamma\beta_{i}+\alpha_{i}\chi\left(\beta_{i}\right)\right]+\sqrt{\chi\left(\beta_{i}\right)}\xi_{i}^{\beta}+f\sum_{j}J_{ij}\xi_{j},\nonumber \\
\,\label{eq:SDE_posP}
\end{eqnarray}
with the definitions that:
\begin{align}
\chi\left(\alpha\right) & \equiv\frac{\kappa}{\gamma_{p}}\left[\epsilon_{p}-\frac{\kappa}{2}\alpha^{2}\right]\,\nonumber \\
\epsilon_{i} & =\zeta\sum_{j}J_{ij}\left(\alpha_{j}+\beta_{j}\right)\,.
\end{align}

The equations above are also It\^{o} stochastic equations, although
the noise terms correspond to the total quantum noise in the system
itself. Stratonovich equations, which are more tractable numerically,
are then obtained by the mapping of $\gamma\rightarrow\gamma'\equiv\gamma-\kappa^{2}/4\gamma_{p}$
\citep{Drummond_OpActa1981}. This allows one to use more robust and
accurate numerical techniques \citep{drummond1991computer}.

\subsection{Conditional phase-space simulations}

In Eq (\ref{eq:SDE_posP}), the measurement noise has been averaged
over, thus removing it from the equations. Because of this, it allows
for a very efficient numerical simulation of the CIM. In addition
to the total master equation given by Eq (\ref{eq:CIM_fb_master_full}),
we would also like to simulate the conditional master equation with
the measurement noise present. We expect that simulating the conditional
master equation for different realizations of the measurement noise
and averaging the simulation outcome will produce results consistent
with those obtained from the total master equation.

While a conditional master equation approach is not the most efficient
one, there are nevertheless good reasons to pursue it. For one, it
allows us to carry out a consistency check between the conditional
and total master equations. At the same time, there might be situations
where (at least partial) knowledge of the measurement noise exists,
for example when the measurement outcome is recorded. Furthermore,
there are cases for which a total master equation might not be found
as easily. One such case is where the finite time delay is to be taken
into account explicitly. In a phase-space simulation, this could mean
applying feedback based on the system state and measurement noise
from one or more time-steps ago. Also, a total master equation may
not be found as easily if the feedback is not strictly proportional
to the measurement outcome. 

In our derivation of the total master equation, we have made the assumption
that the feedback is proportional to the measurement outcome immediately
after the collapse of the wave-function. This enabled us to find Eq
(\ref{eq:fb_total_w_noise}), where terms of second order in $\rho_{c}$
(through the trace operator included in $\mathcal{H}$) only appear
multiplied with the measurement noise. By averaging over measurement
noises, we removed this term and obtained a fully deterministic master
equation which is strictly linear in $\rho$. If the feedback is not
proportional to the measurement outcome $I_{s}\left(t\right)$, averaging
over the measurement noise would most likely result in a master equation
with higher-order terms in $\rho$ (through the trace operator), which
is forbidden\citep{Primas1990}.

We now attempt to formulate a set of phase-space equations with which
to simulate the conditional master equation. The most obvious approach
would be to apply the familiar chain of transformations ``quantum
master equation -> Fokker-Planck equation -> stochastic differential
equations'' to Eq (\ref{eq:fb_total_w_noise}) as we did for Eq (\ref{eq:CIM_fb_master_full}).
However, there are several problems with Eq (\ref{eq:fb_total_w_noise})
when it comes to finding a corresponding Fokker-Planck equation: First,
there is a noise term $\xi$$\left(t\right)$, which means the master
equation itself is a stochastic equation. A Fokker-Planck equation
is a deterministic partial differential equation. Furthermore, a Fokker-Planck
equation has only first- and second-order derivative terms with respect
to its phase-space variables. Eq (\ref{eq:fb_total_w_noise}) would
clearly lead to non-derivative terms due to the operator $\mathcal{H}\left[c_{i}\right]$.
Lastly, due to the expectation value in $\mathcal{H}\left[c_{i}\right]$,
the corresponding equation describing the phase-space distribution
would result in an integro-differential equation, another difference
to a conventional Fokker-Planck equation.

Hush et al.\citep{hush2009scalable} have investigated the question
how a Fokker-Planck like equation with these features can be simulated
efficiently using stochastic samples. They consider a general equation
of the form
\begin{eqnarray}
dP & = & \left\{ \left(-\sum_{i}\partial_{i}A_{i}+\frac{1}{2}\sum_{i,j}\partial_{i}C_{ij}\partial_{i'}C_{i'k}+\iota-\left\langle \iota\right\rangle \right)dt\right.\nonumber \\
 &  & \left.+\sum_{j}\left(-\sum_{i}\partial_{i}B_{ij}+\nu_{j}-\left\langle \nu_{j}\right\rangle \right)dW_{j}^{\left(s\right)}\right\} P,\label{eq:Hush_FPE}
\end{eqnarray}
where $P\equiv P\left(\mathbf{x},\mathbf{dW}^{\left(s\right)}\left(t\right),t\right)$,
while $\mathbf{x}$ and $\mathbf{dW}^{\left(s\right)}$ are vectors
of (phase-space) variables and noise increments, respectively. The
terms $A$, $B$, $C$, $\iota$ and $\nu$ are (vector-, matrix-
and scalar-valued) functions which may depend on $\mathbf{x}$ as
well as the distribution $P$. Unlike the stochastic equations we
have considered so far, Eq (\ref{eq:Hush_FPE}) is understood to be
in the Stratonovich calculus, indicated by the superscript $^{\left(s\right)}$.

Hush et al. demonstrated that Eq (\ref{eq:Hush_FPE}) can be treated
using a set of Stratonovich-type stochastic differential equations
with the addition of a weight variable $\omega\left(t\right)$. The
full set of stochastic equations is:
\begin{eqnarray}
dx_{i} & = & A_{i}dt+\sum_{j}B_{ij}\left(x,t\right)dW_{j}^{\left(s\right)}\nonumber \\
 &  & +\sum_{k}C_{ik}\left(x,t\right)dV_{k}^{\left(s\right)}\label{eq:Hush_dx}\\
\frac{d\omega}{\omega} & = & \iota\left(x,t\right)dt+\sum_{j}\nu_{j}\left(x,t\right)dW_{j}^{\left(s\right)}\,,\label{eq:Hush_dw}
\end{eqnarray}
where both $dW^{\left(s\right)}=dW^{\left(s\right)}\left(t\right)$
and $dV^{\left(s\right)}=dV^{\left(s\right)}\left(t\right)$ are Stratonovich-type
noise increments. However, there is a profound difference between
these two noise terms. The $dV$ terms originate from second-order
derivatives in Eq (\ref{eq:Hush_FPE}) just like for a conventional
Fokker-Planck equation. As such, they are independently drawn for
every stochastic sample that is simulated. In contrast, the $dW$
terms correspond to the measurement noise in Eq (\ref{eq:Hush_FPE})
and is drawn once per time-step for the entire stochastic ensemble.
Due to the nature of the $dV$ terms, they are called ``fictitious''
noises, while the $dW$ terms are called ``real'' noises.

Any observables $f\left(\mathbf{x}\right)$ based on the conditional
equations are obtained via
\begin{eqnarray}
\overline{f\left(\mathbf{x}\right)} & \equiv & \mathbb{E}\left[\omega f\left(\mathbf{x}\right)\right]/\mathbb{E}\left[\omega\right]\,.\label{eq:hush_exp_value}
\end{eqnarray}
Here $\text{\ensuremath{\mathbb{E}\left[\cdot\right]}}$ indicates
the average with respect to stochastic trajectories.

While the above method in principle will provide the correct predictions,
it is obvious from Eq (\ref{eq:Hush_dw}) that the noises are likely
to cause numerical instabilities due exponential decay and growth.
In order to make the simulations more tractable, there are three additions
that can be made to the conventional Monte Carlo simulation algorithm
of Eqs (\ref{eq:Hush_dx}) and (\ref{eq:Hush_dw}).

The first and most important addition is a technique called breeding.
Its purpose is to ``even out'' the distribution of weights by cloning
the highest-weighted trajectories into two copies with half their
original weight while simultaneously removing trajectories
with extremely low weight. More precisely, the breeding algorithm
consists of the following steps:
\begin{enumerate}
\item Find the trajectory index $i$ with lowest weight $\omega_{min}$.
Calculate the ratio $r$ between lowest weight and average weight
$r\leftarrow\omega_{min}/\left\langle \omega\right\rangle $. If $r$
is less than a cutoff ratio $\epsilon_{thr,breed}$, continue with
the next steps. Otherwise, do nothing (terminate).
\item Find the trajectory index $j$ with highest height $\omega_{max}$.
Replace $i$'th trajectory by $j$'th trajectory, i.e. $\mathbf{x}_{i}\leftarrow\mathbf{x}_{j}$.
Set the weights of both trajectories to half of $\omega_{max}$, i.e.
$\omega_{i}\leftarrow\omega_{max}/2$, $\omega_{j}\leftarrow\omega_{max}/2$.
\item Go back to step 1.
\end{enumerate}
We have found that the breeding algorithm works best when executed
after every time-step in the stochastic integration. When running
the stochastic integration, we are recording the number of ``breed''
events, that is, the number of times step 2 is executed.

Another addition which improves numerical stability is to normalize
the weights, i.e. $\omega\leftarrow\omega/\left\langle \omega\right\rangle $.
In our simulations, this is done following the breeding algorithm.

Lastly, instead of simulating the weights $\omega$ themselves, we
are using the transformed weights $\omega'=\log\left(\omega\right)$.
This leads to differential equation for the transformed weights:
\begin{eqnarray}
d\omega' & = & \iota\left(x\left(t\right),t\right)dt+\sum_{j}\nu_{j}\left(x\left(t\right),t\right)dW_{j}^{\left(s\right)}\left(t\right).
\end{eqnarray}

In order to apply the above method to the Coherent Ising machine,
it is necessary to reconsider the issue of feedback, the reason being
that Eq (\ref{eq:Hush_FPE}) is a Stratonovich-type stochastic equation,
whereas so far, we have treated measurement-feedback systems entirely
in the It\^{o} scheme. Note that for measurement-feedback systems,
the choice of integration scheme has a subtle effect on the interpretation
of the feedback noise as will be shown shortly.

After mapping the Stratonovich-type master equation given in Eq (\ref{eq:fb_total_w_noise_Strat})
to the positive-P representation, which results in a Fokker-Planck
like equation, one can apply the weighted integration scheme. This
results in the Stratonovich-type stochastic differential equations
\begin{eqnarray}
\dot{\alpha}_{i} & = & \left[\epsilon_{i}+\left(\frac{\kappa^{2}}{4\gamma_{p}}-\gamma\right)\alpha_{i}^{2}+\beta_{i}\chi\left(\alpha_{i}\right)\right]+\sqrt{\chi\left(\alpha_{i}\right)}\xi_{i}^{\alpha}\nonumber \\
\dot{\beta}_{i} & = & \left[\epsilon_{i}+\left(\frac{\kappa^{2}}{4\gamma_{p}}-\gamma\right)\beta_{i}^{2}+\alpha_{i}\chi\left(\beta_{i}\right)\right]dt+\sqrt{\chi\left(\beta_{i}\right)}\xi_{i}^{\beta}\nonumber \\
\dot{\omega} & = & \gamma_{m}\sum_{i}\left(\alpha_{i}+\beta_{i}\right)\left(2\left\langle \alpha_{i}+\beta_{i}\right\rangle -\left(\alpha_{i}+\beta_{i}\right)\right)\nonumber \\
 &  & +\sqrt{2\gamma_{m}}\sum_{i}\left(\alpha_{i}+\beta_{i}\right)\xi_{i}^{r}\,,
\end{eqnarray}
where
\begin{equation}
\epsilon_{i}=\zeta\sum_{j}J_{ij}\left(\left\langle \alpha_{j}+\beta_{j}\right\rangle +\frac{\xi_{i}^{r}}{\sqrt{2\gamma_{m}}}\right)\,.
\end{equation}
Here, $\xi_{i}^{\alpha},\:\xi_{i}^{\beta}$ correspond to ``fictitious''
noises while $\xi_{i}^{r}$ correspond to measurement (``real'')
noises. 

\section{Numerical results}

Three different types of simulation were carried out to illustrate
and compare the methods. While the coherent state expansion means
that one can only rigorously compare the averages over many trajectories,
in fact the method is even more powerful than this. Since the final
state has a macroscopic distinction between ``spin-up'' and ``spin-down'',
one can also compare the actual distributions of the final results,
and this will correspond to the corresponding experimental distributions
due to their macroscopicity.

\subsection{Small-scale pump ramps}

An experiment is considered with $N=16$ degenerate parametric oscillators.
The interaction matrix $\mathbf{J}$ corresponds to the 1-dimensional
(circular) antisymmetric Ising model, that is $J_{ij}=-1$ if $\left|i-j\right|=1$
or $\left|i-j\right|=N-1$, $J_{ij}=0$ otherwise.

The system parameters are $\gamma_{s}=1.0$, $\gamma_{m}=0.1$, $\gamma_{p}=10$,
$\kappa=0.1$. $N_{T}=500\cdot10^{3}$ time-steps and $N_{s}=8192$
stochastic samples were used for the integration. 

The pump strength was linearly increased from $\varepsilon_{p}=0$
to $\varepsilon_{p}=2\cdot\varepsilon_{p,th}$ with $\varepsilon_{p,th}=\frac{\gamma\gamma_{p}}{\kappa}$,
where $\gamma\equiv\gamma_{s}+\gamma_{m}$ during the integration
time.

The integration was carried out for the total master equation as well
as the conditional master equation with the weighted scheme explained
previously using a stochastic RK4 integration scheme. For the weighted
scheme, the weight rebalancing (breeding) algorithm was carried out
after each time-step using a breeding threshold of $\epsilon_{th}=10^{-4}$.

The simulation was repeated for 3 different integration times and
9 different values for the feedback parameter $\zeta$. 

\begin{figure}
\includegraphics[width=0.8\columnwidth]{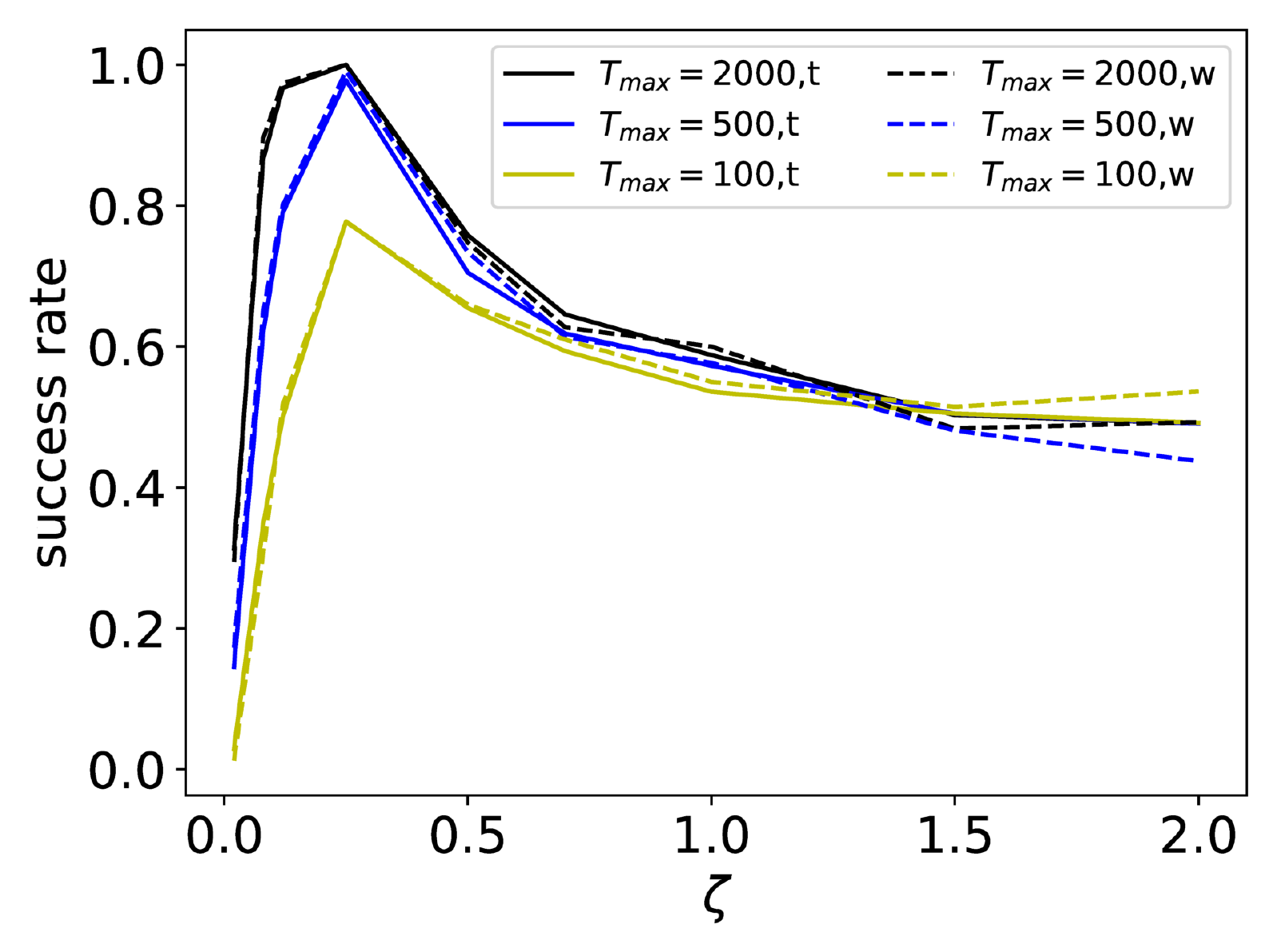}

\caption{Success probabilities for the CIM with $N=16$ sites and linearly
increased pump field, as a function of the (constant) feedback strength
$\zeta$ for three different integration times. The results were obtained
using the conditional (weighted) integration scheme (abbreviated with
``w'' in the legend) and the total master equation method (abbreviated
with ``t'' in the legend).}
\label{fig:succ_prob_N16}
\end{figure}

A success rate is defined as the fraction of instances for which the
simulated system ascertains the Ising model ground state. Here, the
ground state is given by the degenerate states $\left(+,-,...,+,-\right)$
and $\left(-,+,...,-,+\right)$, where the spin states are given by
the sign of the mode's x-quadrature. In the case of the total master
equation, the success rate can be calculated by considering all $N_{s}$
individual trajectories. Here, $20$ independent simulations with
$N_{s}$ trajectories were used to determine the error of the mean,
which is negligibly small. For the case of the weighted simulations,
this is considerably more resource intensive, demonstrating the clear
superiority of the total master equation method. Here, the entire
stochastic ensemble is needed to determine a mode's x-quadrature according
to Eq (\ref{eq:hush_exp_value}). The experiment is repeated $200$
times in order to determine the success probability. However, there
is not enough data to determine the error of the mean. The results are shown in Fig. \ref{fig:succ_prob_N16}.

\begin{figure}
\includegraphics[width=0.8\columnwidth]{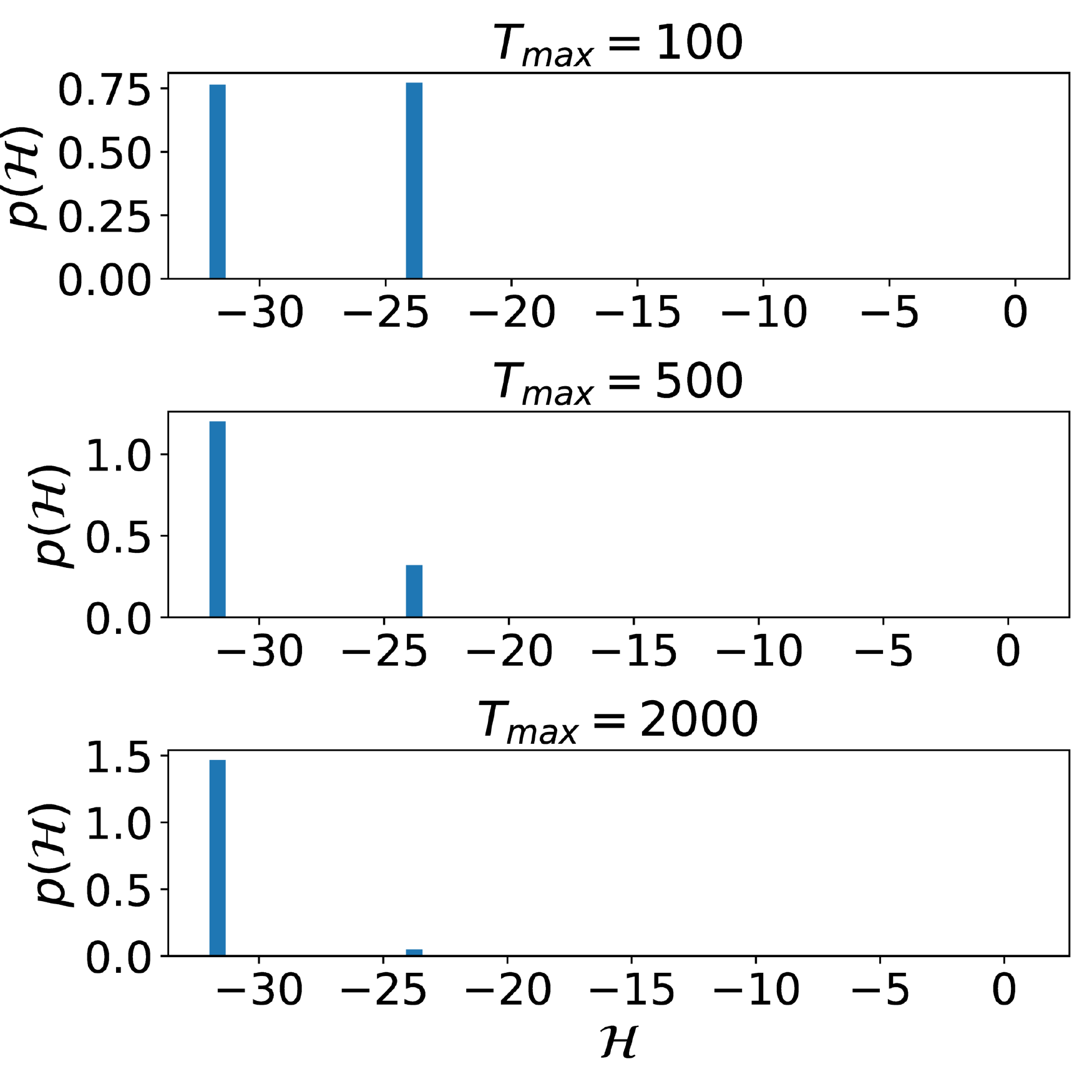}

\caption{Probability density of the Ising model Hamiltonian for the CIM with
$N=16$ sites and linearly increased pump field with a constant feedback
strength of $\zeta=0.12$ for three different integration times. The
results were obtained using the total master equation method and are
based on $6,144$ stochastic trajectories.}
\label{fig:Prob_dens_N16}
\end{figure}

In the case of the total master equation for an interaction strength
of $\zeta=0.12$, the probability density for the Ising model Hamiltonian
is recorded, which is defined as
\begin{eqnarray}
\mathcal{H} & = & -\sum_{i,j}s_{i}\left(\mathbf{J}_{ij}\right)s_{j}\,,
\end{eqnarray}
where $s_{i}$ indicates the corresponding spin state given by the
$x$-quadrature of the i'th DPO mode, that is
\begin{eqnarray}
s_{i} & \equiv & sgn\left(\Re\left(\alpha_{i}+\beta_{i}\right)\right)\,.
\end{eqnarray}
The probability density is shown in Fig. \ref{fig:Prob_dens_N16}.

\subsection{Small-scale pump and feedback ramps }

\begin{figure}
\includegraphics[width=0.8\columnwidth]{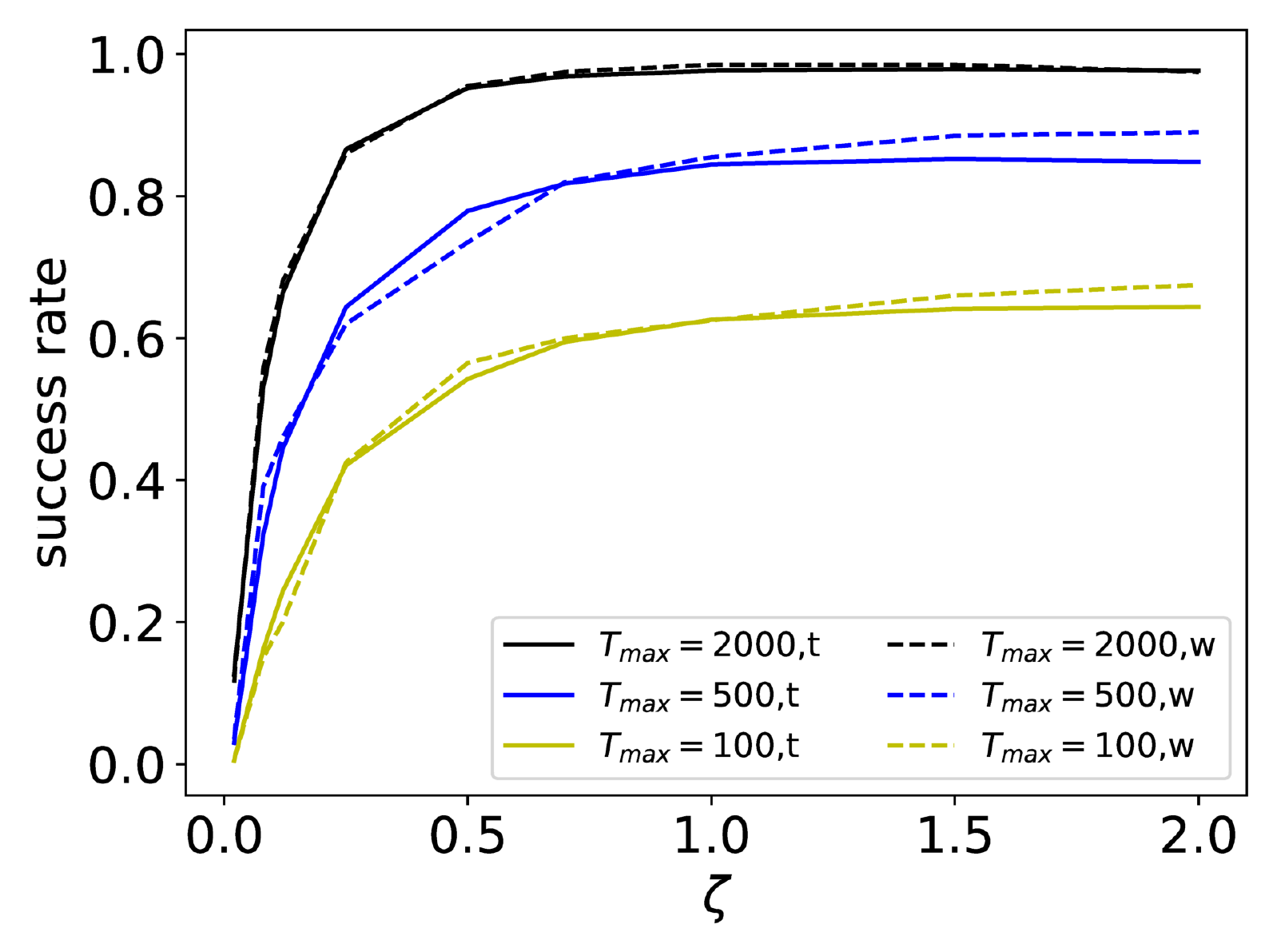} \caption{Success probabilities for the CIM with $N=16$ sites and linearly
increased pump field and feedback strength, as a function of the maximum
feedback strength $\zeta_{max}$ for three different integration times.
The results were obtained using both the conditional (weighted) integration
scheme (abbreviated with ``w'' in the legend) and the total master
equation method (abbreviated with ``t'' in the legend).}
\label{fig:succ_prob_N16_mod}
\end{figure}

A second experiment is considered with the system parameters given
above. Here, a different minimization strategy is employed. Where
in the first experiment, the pump strength was linearly increased
during the simulation time, now the pump strength as well as the feedback
parameter $\zeta$ are linearly increased to $2\cdot\varepsilon_{p,th}$
and $\zeta=\zeta_{max}$, respectively. As with the first experiment,
$20$ independent simulations were used in the case of the total master
equation simulations to estimate the error of the mean, while $200$
independent repetitions to estimate the success probability in the
case of the conditional (weighted) method. This is shown in Fig. \ref{fig:succ_prob_N16_mod}. 

We note the greatly improved success rate at large feedback strengths,
indicating the sensitivity of the CIM to different ramp strategies.
As before, there is excellent agreement between the conditional and
unconditional methods.

\subsection{Large-scale pump ramps }

Lastly, we consider an experiment consisting of $N=1,000$ parametric
oscillators, using a linear pump ramp. Due to its large size, only
the much faster unconditional simulations were run.

As outlined in Sec. \ref{subsec:The-Ising-model}, an Ising model
can be identified with a (weighted, undirected) graph with the interaction
matrix $\mathbf{J}$ corresponding to the adjacency matrix and weight
function, respectively. Here, the interaction matrix is chosen so
that it corresponds to a random graph generated using the following
set of rules:
\begin{itemize}
\item Each of the $\frac{N\cdot\left(N-1\right)}{2}$ edges has non-zero
weight with a probability of $p$ and zero weight with a probability
of $1-p$. Here, $p=0.1$.
\item A non-zero weight is either $+1$ or $-1$ with equal probability.
\end{itemize}
In other words, $\mathbf{J}$ is a symmetric $1,000\times1,000$ matrix
with main diagonal entries equal to zero off-diagonal entries equal
to $0$ with probability $1-p$, $+1$ with probability $\frac{p}{2}$
and $-1$ with probability $\frac{p}{2}$.

The interaction matrix is generated once and used for all quantum
trajectories of the stochastic simulation.

The system parameters are $\gamma_{s}=1.0$, $\gamma_{m}=0.1$, $\gamma_{p}=10$,
$\kappa=0.1$, $N_{T}=10\cdot10^{3}$, $\zeta=0.1$ and $T_{max}=10$.
The pump strength is linearly increased from $\varepsilon_{p}=0$
to $\varepsilon_{p}=3\cdot\varepsilon_{p,th}$ during the integration
time. Due to the exponential complexity of the problem, precise knowledge
of the ground state of the system considered here is likely impossible.
Hence, instead of the success probability for finding the ground state,
the Ising model interaction Hamiltonian is considered.

\begin{figure}
\includegraphics[width=0.8\columnwidth]{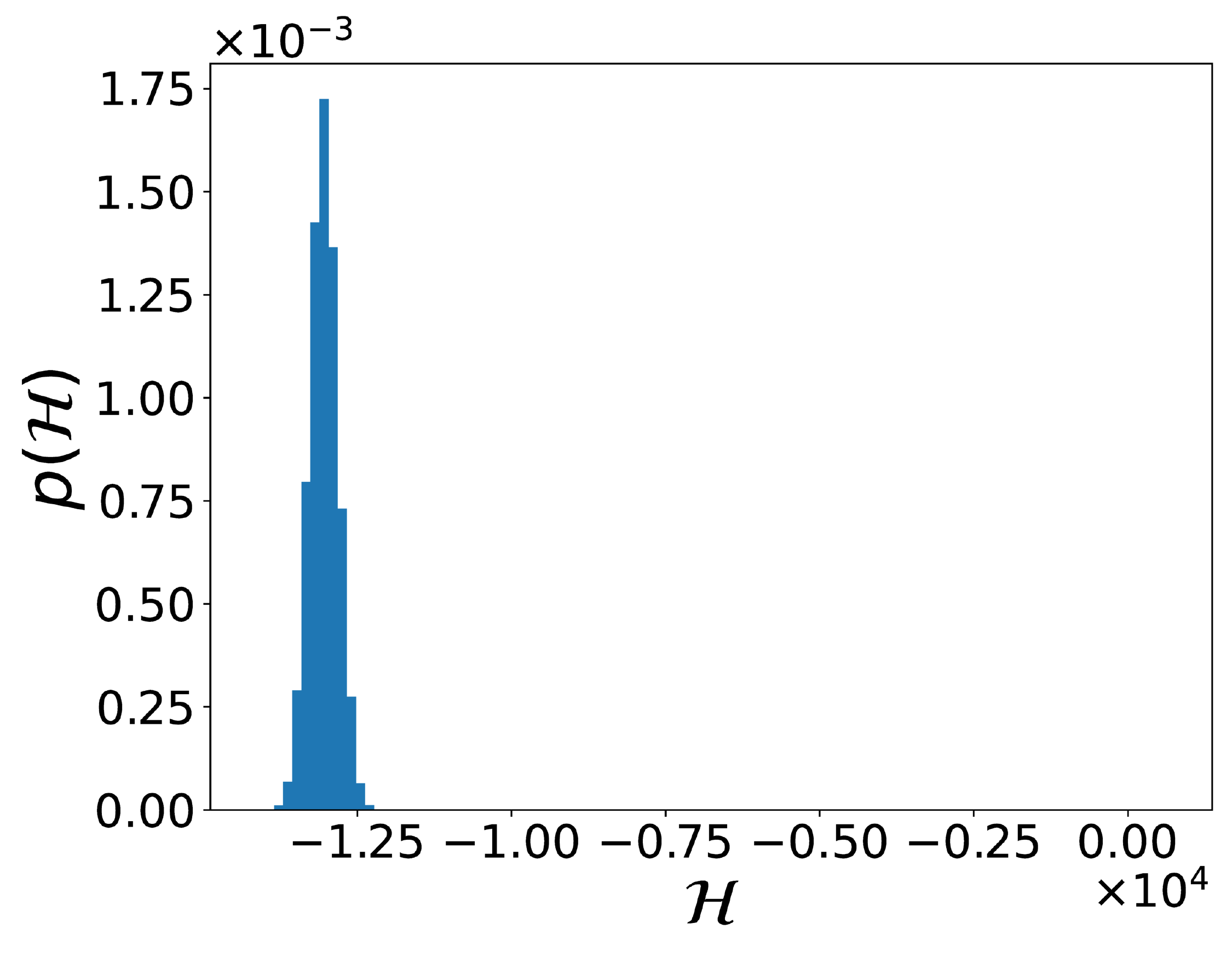}

\caption{Probability distribution for the outcome of the Ising model interaction
Hamiltonian for a graph consisting of $1,000$ nodes with a connectivity
of $10\%$.}
\label{fig:prob_dens_1000_nodes}
\end{figure}

\begin{figure}
\includegraphics[width=0.8\columnwidth]{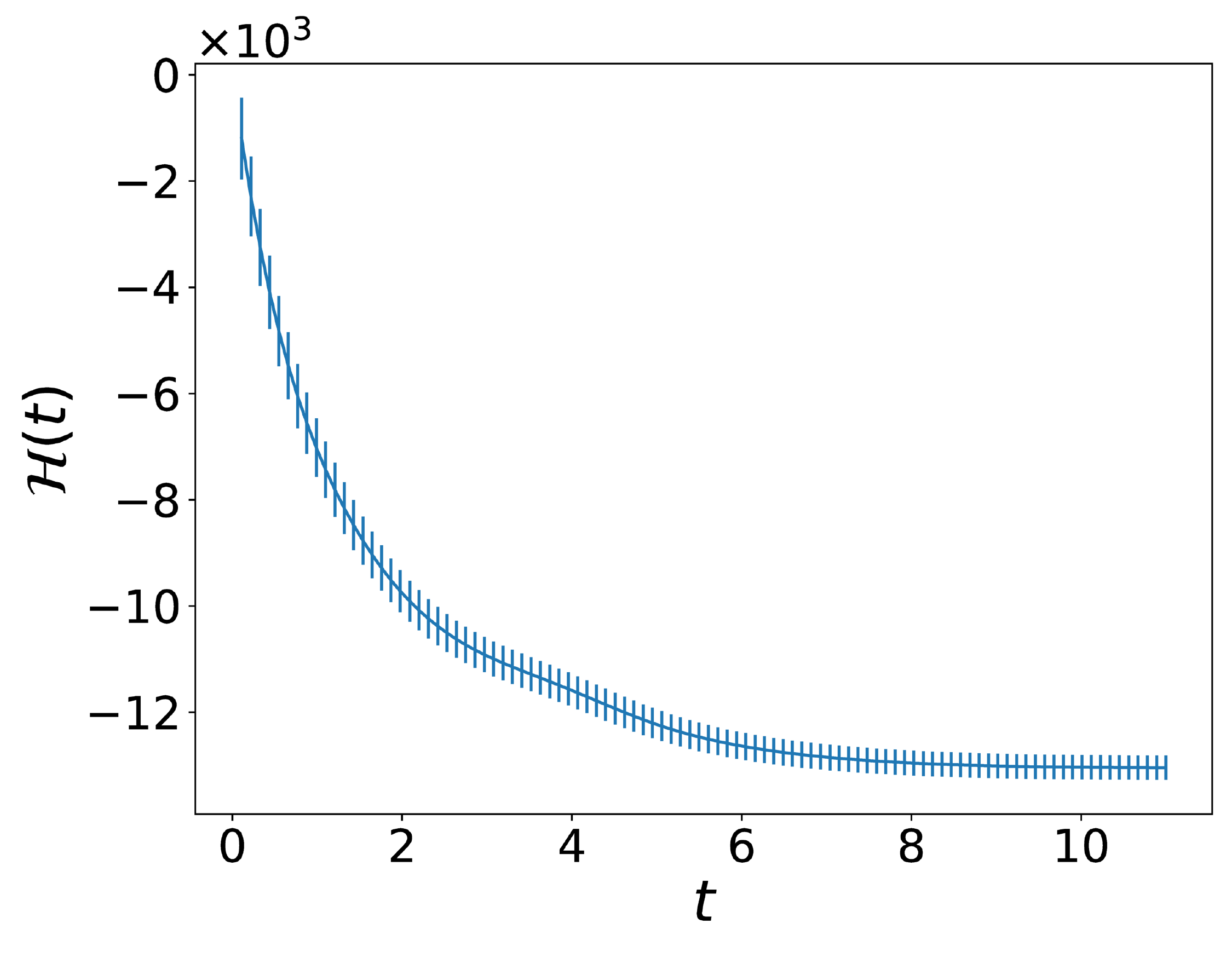}

\caption{Evolution of the mean Ising model interaction Hamiltonian for a graph
consisting of $1,000$ nodes as a function of simulation time. Here,
vertical lines indicate the standard deviation, obtained using $N_{S}=102,400$
stochastic trajectories.}
\label{fig:evol_1000_nodes}
\end{figure}

The simulation was carried out using a total of $N_{S}=102,400$ stochastic
trajectories with interaction strength of $\zeta=0.05$. They took
ca. 12 hours on 40 state-of-the-art GPU's running in parallel on a
computer cluster. The resulting probability density for a given outcome of the
interaction energy is given in Fig. \ref{fig:prob_dens_1000_nodes}, while Fig. \ref{fig:evol_1000_nodes} shows the evolution of the
mean interaction energy as a function of simulation time.

Since our purpose here to demonstrate the scalability of the unconditional
method, we did not optimize the ramp strategy. However, as can be
seen above, it is likely that an optimized ramp strategy would give
a narrower distribution and a higher success rate, which is clearly
desirable in finding the best solution.

\section{Conclusion}

The Coherent Ising Machine is a promising, novel technology with potential
applications in a number of areas. An advantage over classical computers
for a specific problem has been claimed \citep{honjo2021100}, although
the role of quantum effects remains unclear. Unlike many candidates
for gate array-type quantum computers, the CIM can be operated at
room temperature and has very stable states. For the measurement-feedback
architecture, the size of the CIM can be scaled up quite easily. While
it is likely that quantum effects only play a minor role if at all
in contemporary realizations of the CIM, entering a regime for which
these become relevant could possibly lead to even better performance
due to quantum tunneling and other effects.

In such a regime, precise and reliable simulation methods are required
to better understand the role of quantum effects. We have derived
two different methods, both utilizing the generally non-approximate
positive-P phase-space representation. The conditional, weighted method
allows for the simulation of a single instance or run of a given measurement-feedback
CIM. It fully ``captures'' the effect of the partial state collapse
induced by the homodyne measurement process. The total master equation
allows for the simulation of an average over a large number of independent
runs of the CIM. Here, the state collapse operator is removed from
the master equation through the averaging process. Though the two
methods are quite different in the way they have been derived and
in terms of their numerical implementation, we have demonstrated that
they are in good agreement for the CIM considered here.

The latter method is several orders of magnitude faster compared to
the conditional one. This comes at the price of being limited to predictions
for an overage over multiple runs of the CIM. However, one would most
likely be interested in the outcome of a CIM averaged over multiple
runs anyway, making the total master equation method the more useful
choice due to its significantly lower computational demand.
\begin{acknowledgments}
This work was partly performed on the OzSTAR national facility at
Swinburne University of Technology. OzSTAR is funded by Swinburne
University of Technology and the National Collaborative Research Infrastructure
Strategy (NCRIS). This research was funded through a grant from NTT
Phi Laboratories.
\end{acknowledgments}

\section*{Appendix}

Here we derive the form of the Stratonovich corrections for the feedback
master equation. In general there are two terms, from measurement
and from feedback. 

\subsection*{Measurement}

In the case of measurement, the operator $\mathcal{H}\left[c\right]\rho_{c}$
includes a mean value term, which means that it is nonlinear in the
density matrix components. This leads to the additional terms described
here. We only treat the single-mode case in this Appendix, as the
multi-mode case is similar. 

Writing out the B-matrix for the single-mode case, and ignoring the
$k$ index since $k=1$, we note that
\begin{eqnarray}
B & =\left\{ \mathcal{H}\left[c\right]\rho_{c}\right\} = & c\rho_{c}+\rho_{c}c^{\dagger}-T\rho_{c},
\end{eqnarray}
withe the definition that $T\equiv\sum_{kl}\left[c_{lk}+c_{lk}^{\dagger}\right]\rho_{kl}\equiv\sum_{kl}t_{lk}\rho_{kl}=\left\langle c+c^{\dagger}\right\rangle _{c}$,
provided that $\sum_{k}\rho_{kk}=1$.

We now use an orthonormal basis expansion to transform the conditional
density operator $\rho_{c}$ into a matrix $\rho_{mn}$. The matrix
derivative of the $B$ matrix with respect to $\rho_{mn}$ is given
by:
\begin{eqnarray}
\frac{\partial B_{ij}}{\partial\rho_{mn}} & = & \frac{\partial}{\partial\rho_{mn}}\left\{ c_{ik}\rho_{kj}+\rho_{ik}c_{kj}^{\dagger}-\rho_{ij}\sum_{kl}t_{lk}\rho_{kl}\right\} \nonumber \\
 & = & \left\{ c_{im}\delta_{jn}+\delta_{im}c_{nj}^{\dagger}-\delta_{im}\delta_{jn}T-\rho_{ij}t_{nm}\right\} .
\end{eqnarray}
Therefore we see immediately that the Stratonovich correction term
$\text{C}^{\mathcal{H}}$ is given by:
\begin{align}
C_{ij}^{\mathcal{H}} & =-\frac{1}{2}\sum_{mn}B_{mn}\frac{\partial}{\partial\rho_{mn}}B_{ij}\nonumber \\
 & =-\frac{1}{2}\sum_{mn}\left\{ c_{mk}\rho_{kn}+\rho_{mk}c_{kn}^{\dagger}-\rho_{mn}T\right\} \times\nonumber \\
 & \,\,\,\,\,\,\,\,\,\left\{ c_{im}\delta_{jn}+\delta_{im}c_{nj}^{\dagger}-\delta_{im}\delta_{jn}T-\rho_{ij}t_{nm}\right\} \,.
\end{align}

There are 12 terms in this product, and they are listed below: 
\begin{enumerate}
\item $c_{mk}\rho_{kn}c_{im}\delta_{jn}=c_{im}c_{mk}\rho_{kj}=\left(cc\rho_{c}\right)_{ij}$
\item $c_{mk}\rho_{kn}\delta_{im}c_{nj}^{\dagger}=c_{ik}\rho_{kn}c_{nj}^{\dagger}=\left(c\rho_{c}c^{\dagger}\right)_{ij}$
\item $-c_{mk}\rho_{kn}\delta_{im}\delta_{jn}T=-c_{ik}\rho_{kj}T=-\left(c\rho_{c}T\right)_{ij}$
\item $-c_{mk}\rho_{kn}\rho_{ij}t_{nm}=-\rho_{ij}Tr\left(c\rho_{c}t\right)$
\item $\rho_{mk}c_{kn}^{\dagger}c_{im}\delta_{jn}=c_{im}\rho_{mk}c_{kj}^{\dagger}=\left(c\rho_{c}c^{\dagger}\right)_{ij}$
\item $\rho_{mk}c_{kn}^{\dagger}\delta_{im}c_{nj}^{\dagger}=c_{kn}^{\dagger}\rho_{ik}c_{nj}^{\dagger}=\left(\rho_{c}c^{\dagger}c^{\dagger}\right)_{ij}$
\item $-\rho_{mk}c_{kn}^{\dagger}\delta_{im}\delta_{jn}T=-c_{kn}^{\dagger}\rho_{ik}T=-\left(\rho_{c}c^{\dagger}T\right)_{ij}$
\item $-\rho_{mk}c_{kn}^{\dagger}\rho_{ij}t_{nm}=-\rho_{ij}Tr\left(t\rho_{c}c^{\dagger}\right)$
\item $-\rho_{mn}Tc_{im}\delta_{jn}=-c_{im}T\rho_{mj}=-\left(Tc\rho_{c}\right)_{ij}$
\item $-\rho_{mn}T\delta_{im}c_{nj}^{\dagger}=-T\rho_{in}c_{nj}^{\dagger}=-\left(T\rho_{c}c^{\dagger}\right)_{ij}$
\item $\rho_{mn}T\delta_{im}\delta_{jn}T=\rho_{ij}T^{2}$
\item $\rho_{mn}T\rho_{ij}t_{nm}=\rho_{ij}T\left(\rho_{mn}t_{nm}\right)=T\rho_{ij}Tr\left(\rho_{c}t\right)$
\end{enumerate}
Combining all these, and returning to an index-free operator notation,
we obtain that: 
\begin{align}
\text{C}^{\mathcal{H}} & =-\frac{1}{2}\left[cc\rho_{c}+2c\rho_{c}c^{\dagger}+\rho_{c}c^{\dagger}c^{\dagger}\right.\nonumber \\
 & -2T\left(c\rho_{c}+\rho_{c}c^{\dagger}\right)+\nonumber \\
 & \left.+\rho_{c}T^{2}+\rho_{c}Tr\left(T\rho_{c}t-c\rho_{c}t-t\rho_{c}c\right)\right].
\end{align}

The last term with factors of $T$ is: 
\begin{equation}
T^{2}+Tr\left(T\rho_{c}t-c\rho_{c}t-t\rho_{c}c^{\dagger}\right)=2T^{2}-\left\langle c^{2}+c^{\dagger2}+2c^{\dagger}c\right\rangle .
\end{equation}
Combining terms together, one obtains that:
\begin{equation}
\text{C}^{\mathcal{H}}=\left\langle c+c^{\dagger}\right\rangle _{c}\mathcal{H}\left[c\right]\rho_{c}-c\rho_{c}c^{\dagger}+\rho_{c}\left\langle c^{\dagger}c\right\rangle -\frac{1}{2}\mathcal{H}\left[cc\right]\rho_{c}.\label{eq:C_H}
\end{equation}

\subsection*{Measurement and feedback}

In the case of measurement with feedback, the operator multiplying
the noise is $\xi\left(t\right)\left(\mathcal{H}+\mathcal{K}\right)\rho_{c}$,
where we will assume that $\mathcal{K}\rho_{c}\equiv\left[K,\rho_{c}\right]$
. Writing out the $B$-matrix for the single-mode case, we notice
that
\begin{eqnarray}
B & = & B^{\mathcal{H}}+B^{\mathcal{K}}\nonumber \\
 & = & c\rho_{c}+\rho_{c}c^{\dagger}-T\rho_{c}+K\rho_{c}-\rho_{c}K.
\end{eqnarray}
The matrix derivative is:
\begin{eqnarray}
\frac{\partial B_{ij}}{\partial\rho_{mn}} & = & \frac{\partial}{\partial\rho_{mn}}\left\{ B_{ij}^{\mathcal{H}}+\left[K\rho_{c}-\rho_{c}K\right]_{ij}\right\} \nonumber \\
 & = & \left\{ B_{ij,mn}^{\mathcal{H}}+K_{im}\delta_{jn}-\delta_{im}K_{nj}\right\} .
\end{eqnarray}

As a result, the total Stratonovich correction is:
\[
C_{ij}=C_{ij}^{\mathcal{H}}+C_{ij}^{\mathcal{HK}}+C_{ij}^{\mathcal{KH}}+C_{ij}^{\mathcal{K}}
\]
where $C^{\mathcal{H}}$ was obtained already, and we obtain 
\begin{align}
C_{ij}^{\mathcal{HK}} & =-\frac{1}{2}\sum_{mn}B_{mn}^{\mathcal{H}}\frac{\partial}{\partial\rho_{mn}}B_{ij}^{\mathcal{K}}\nonumber \\
 & =-\frac{1}{2}\sum_{mn}\left\{ c_{mk}\rho_{kn}+\rho_{mk}c_{kn}^{\dagger}-\rho_{mn}T\right\} \times\nonumber \\
 & \,\,\,\,\,\,\,\,\,\left\{ K_{im}\delta_{jn}-\delta_{im}K_{nj}\right\} \nonumber \\
 & =-\frac{1}{2}\left[K,c\rho_{c}+\rho_{c}c^{\dagger}-\rho_{c}T\right]_{ij}\,,\label{eq:C_HK}
\end{align}

\begin{align}
C_{ij}^{\mathcal{KH}} & =-\frac{1}{2}\sum_{mn}B_{mn}^{\mathcal{K}}\frac{\partial}{\partial\rho_{mn}}B_{ij}^{\mathcal{H}}\nonumber \\
 & =-\frac{1}{2}\sum_{mn}\left[K\rho_{c}-\rho_{c}K\right]_{mn}\times\nonumber \\
 & \,\,\,\,\,\,\,\,\,\left\{ c_{im}\delta_{jn}+\delta_{im}c_{nj}^{\dagger}-\delta_{im}\delta_{jn}T-\rho_{ij}t_{nm}\right\} \nonumber \\
 & =-\frac{1}{2}\left[c\left[K,\rho_{c}\right]+\left[K,\rho_{c}\right]c^{\dagger}\right.\nonumber \\
 & \:\,\,\,\,\,\,\,\,\,\left.-T\left[K,\rho_{c}\right]-\rho_{c}Tr\left[\left[K,\rho_{c}\right]t\right]\right]_{ij},\label{eq:C_KH}
\end{align}

\begin{align}
C_{ij}^{\mathcal{K}} & =-\frac{1}{2}\sum_{mn}B_{mn}^{\mathcal{K}}\frac{\partial}{\partial\rho_{mn}}B_{ij}^{\mathcal{K}}\nonumber \\
 & =-\frac{1}{2}\sum_{mn}\left[K,\rho_{c}\right]_{mn}\times\nonumber \\
 & \,\,\,\,\,\,\,\,\,\left\{ K_{im}\delta_{jn}-\delta_{im}K_{nj}\right\} \nonumber \\
 & =-\frac{1}{2}\left[K,\left[K,\rho_{c}\right]\right]_{ij}.\label{eq:C_K}
\end{align}
Note that if $\left[c,K\right]=Q_{1}$ and $\left[c^{\dagger},K\right]=Q_{2}$
with $Q_{1},Q_{2}\in\mathbb{C}$, it follows that
\begin{eqnarray}
C_{ij}^{\mathcal{KH}} & = & -\frac{1}{2}\left[Kc\rho_{c}-c\rho_{c}K+K\rho_{c}c^{\dagger}-\rho_{c}c^{\dagger}K+\left(Q_{1}+Q_{2}\right)\rho_{c}\right.\nonumber \\
 &  & \:\,\,\,\,\,\,\,\,\,\left.-T\left[K,\rho_{c}\right]-\rho_{c}Tr\left[\rho_{c}\left(Q_{1}+Q_{2}\right)\right]\right]_{ij}\nonumber \\
 & = & C_{ij}^{\mathcal{HK}}\,.
\end{eqnarray}

We assume that these restrictions hold below.

\subsection*{Stratonovich master equation}

Combining the terms given above, the Stratonovich-type master equation
is:
\begin{eqnarray}
\dot{\rho}_{c}^{Strat} & = & \mathcal{L}\rho_{c}+\mathcal{K}\left(c\rho_{c}+\rho_{c}c^{\dagger}\right)+\frac{1}{2}\mathcal{K}^{2}\rho_{c}\nonumber \\
 &  & +\text{C}^{\mathcal{H}}\rho_{c}+2C^{\mathcal{HK}}\rho_{c}+C^{\mathcal{K}}\rho_{c}\nonumber \\
 &  & +\xi\left(t\right)\circ\left[\mathcal{H}\left[c\right]+\mathcal{K}\right]\rho_{c},
\end{eqnarray}
where the relevant corrections are
\begin{eqnarray}
\text{C}^{\mathcal{H}}\rho_{c} & = & \left\langle c+c^{\dagger}\right\rangle _{c}\mathcal{H}\left[c\right]\rho_{c}-c\rho_{c}c^{\dagger}+\rho_{c}\left\langle c^{\dagger}c\right\rangle -\frac{1}{2}\mathcal{H}\left[cc\right]\rho_{c}\nonumber \\
C^{\mathcal{HK}}\rho_{c} & = & -\frac{1}{2}\left[K,c\rho_{c}+\rho_{c}c^{\dagger}-\rho_{c}T\right]\nonumber \\
C^{\mathcal{K}}\rho_{c} & = & -\frac{1}{2}\left[K,\left[K,\rho_{c}\right]\right].
\end{eqnarray}

Since we are using the operators 
\begin{eqnarray}
c & = & \sqrt{2\gamma_{m}}a\\
K & = & \frac{\zeta}{\sqrt{2\gamma_{m}}}\left(a^{\dagger}-a\right)\,,
\end{eqnarray}
clearly $\left[c,K\right]=Q_{1}$ and $\left[c^{\dagger},K\right]=Q_{2}$
with $Q_{1},Q_{2}\in\mathbb{C}$ are satisfied, from which it follows
that $C^{\mathcal{HK}}=C^{\mathcal{KH}}$, hence
\begin{eqnarray}
\dot{\rho}_{c}^{Strat} & = & \mathcal{L}\rho_{c}+\mathcal{K}\left(c\rho_{c}+\rho_{c}c^{\dagger}\right)+\frac{1}{2}\mathcal{K}^{2}\rho_{c}\nonumber \\
 &  & +\text{C}^{\mathcal{H}}\rho_{c}+2C^{\mathcal{HK}}\rho_{c}+C^{\mathcal{K}}\rho_{c}\nonumber \\
 &  & +\xi\left(t\right)\circ\left[\mathcal{H}\left[c\right]+\mathcal{K}\right]\rho_{c}\nonumber \\
 & = & \mathcal{L}\rho_{c}+T\left[K,\rho_{c}\right]\nonumber \\
 &  & +\text{C}^{\mathcal{H}}\rho_{c}\nonumber \\
 &  & +\xi\left(t\right)\circ\left[\mathcal{H}\left[c\right]+\mathcal{K}\right]\rho_{c}\,.
\end{eqnarray}

From this, we finally obtain that

\begin{align}
\dot{\rho}_{c}^{Strat} & =\mathcal{L}\rho_{c}+c\rho_{c}c^{\dagger}-\rho_{c}\left\langle c^{\dagger}c\right\rangle -\frac{1}{2}\mathcal{H}\left[cc\right]\rho_{c}\nonumber \\
 & +I_{c}\left(t\right)\circ\left[\mathcal{H}\left[c\right]+\mathcal{K}\right]\rho_{c},
\end{align}
where: 
\[
I_{c}\left(t\right)=\xi\left(t\right)+Tr\left[\rho_{c}\left(c+c^{\dagger}\right)\right]\,.
\]

\subsection*{Multi-mode case}

Above results can be generalized for a system comprised of $N$ modes.
In this case, there are $N$ independent measurement noises $\xi_{i}\left(t\right)$.
The $B$-matrices are
\begin{eqnarray}
B_{r}^{\mathcal{H}} & = & c_{r}\rho+\rho c_{r}^{\dagger}-T_{r}\rho\nonumber \\
B_{r}^{\mathcal{K}} & = & \sum_{s}J_{rs}\left(K_{s}\rho_{c}-\rho_{c}K_{s}\right),
\end{eqnarray}
with $T_{r}\equiv\sum_{kl}\left[c_{r;lk}+c_{r;lk}^{\dagger}\right]\rho_{r;kl}\equiv\sum_{kl}t_{r;lk}\rho_{r;kl}=\left\langle c_{r}+c_{r}^{\dagger}\right\rangle $.
Note that in the multi-mode case, the operators are 3-tensors with
the first index indicating the mode. The operators satisfy $c_{s;ik}\rho_{r;kj}=\rho_{r;ij}$
and $K_{s;ik}\rho_{r;kj}=\rho_{r;ij}$ which simplifies the calculations. 

Analogously to the one-mode case, one finds
\begin{eqnarray}
\text{C}^{\mathcal{H}} & = & \sum_{r}\left\langle c_{r}+c_{r}^{\dagger}\right\rangle \mathcal{H}\left[c_{r}\right]\rho_{c}+c_{r}\rho_{c}c_{r}^{\dagger}-\rho_{c}\left\langle c_{r}^{\dagger}c_{r}\right\rangle -\frac{1}{2}\mathcal{H}\left[c_{r}c_{r}\right]\rho_{c}\nonumber \\
\nonumber \\
C^{\mathcal{HK}} & = & -1\sum_{rs}J_{rs}\left[K_{s},c_{r}\rho_{c}+\rho_{c}c_{r}^{\dagger}-\rho_{c}T_{r}\right]\nonumber \\
C^{\mathcal{KH}} & = & -\frac{1}{2}\sum_{rs}\left[c_{r}\left[K_{s},\rho_{c}\right]+\left[K_{s},\rho_{c}\right]c_{r}^{\dagger}\right.\nonumber \\
 &  & \:\,\,\,\,\,\,\,\,\,\left.-T_{r}\left[K_{s},\rho_{c}\right]-\rho_{c}Tr\left[\left[K_{s},\rho_{c}\right]t_{r}\right]\right]\nonumber \\
C^{\mathcal{K}} & = & -\frac{1}{2}\sum_{rst}J_{rs}J_{rt}\left[K_{r},\left[K_{t},\rho_{c}\right]\right].
\end{eqnarray}

\bibliographystyle{apsrev4-2}
\bibliography{CIM}

\end{document}